\documentclass[prxquantum,twocolumn,superscriptaddress,longbibliography,nofootinbib,floatfix]{revtex4-2}
\pdfoutput=1
\usepackage[utf8]{inputenc}
\usepackage[english]{babel}
\AtBeginDocument{\renewcommand{\selectlanguage}[1]{}}  
\usepackage[T1]{fontenc}
\usepackage{xcolor}

\usepackage{subcaption}

\DeclareCaptionJustification{apsjustified}{\leftskip=0pt\rightskip=0pt\parfillskip=0pt plus 1fil\relax}
\newcommand{\rl}{\raggedright\let\\\tabularnewline}  
\usepackage{physics}
\usepackage[normalem]{ulem}
\usepackage{xcolor}
\usepackage{soul}
\setstcolor{blue}
\usepackage{float}

\definecolor{detector}{HTML}{87CEEB}   
\definecolor{errorcol}{HTML}{BA55D3}   
\definecolor{logical}{HTML}{32CD32}    
\newcommand{\nD}{\mathrm{D}}
\newcommand{\nEM}{\mathrm{EM}}
\newcommand{\nL}{\mathrm{L}}
\newcommand{\nMP}{N_\mathrm{MP}}
\definecolor{gnnblue}{HTML}{1F77B4}
\definecolor{gnnorange}{HTML}{FF7F0E}
\definecolor{gnngray}{HTML}{7F7F7F}

\usepackage{amsmath}
\usepackage{dsfont}
\usepackage{graphicx}
\usepackage{amssymb}
\usepackage{hyperref}
\hypersetup{
    colorlinks=true,
    linkcolor=logical,
    citecolor=detector,
    urlcolor=errorcol
}
\usepackage{tikz}
\usetikzlibrary{arrows.meta, positioning, calc, fit, decorations.pathreplacing, backgrounds, shadows}
\usepackage{listings}
\usepackage{lipsum}
\usepackage{comment}
\usepackage{multirow}

\lstdefinestyle{python}{
    language=Python,
    basicstyle=\ttfamily\footnotesize,
    keywordstyle=\color{gnnblue}\bfseries,
    commentstyle=\color{gnngray}\itshape,
    stringstyle=\color{gnnorange},
    breaklines=true,
    frame=single,
    framerule=0.4pt,
    rulecolor=\color{gnngray!50},
    backgroundcolor=\color{gnngray!8},
    xleftmargin=4pt,
    xrightmargin=4pt,
    aboveskip=6pt, belowskip=4pt,
}

\begin{document}

\title{A Code-Agnostic Graph Neural Network Decoder from the Detection Error Model}

\author{Federico Alberto Astolfi}
\email{astolfi@unistra.fr}
\affiliation{University of Strasbourg and CNRS, CESQ and ISIS, 67000 Strasbourg, France}

\author{Guido Pupillo}
\email{pupillo@unistra.fr}
\affiliation{University of Strasbourg and CNRS, CESQ and ISIS, 67000 Strasbourg, France}
\affiliation{QPerfect SAS, 23 rue du Loess, Strasbourg, France}
\affiliation{Institut Universitaire de France (IUF), 75000 Paris, France}

\begin{abstract}

We present POLYMECHANON, a graph neural network (GNN) decoder for quantum error correction whose only input is the detection error model (DEM) of a quantum code under a given noise model. We represent the DEM as a tripartite graph built out of detectors, error mechanisms and logical observables, where every input feature is computed by using the quantum code as data rather than design choice. In this way, the same architecture decodes in principle any stabiliser code, under any noise model that can be expressed as a detection error model, once trained on the DEM.
We test this approach along four directions.
Firstly, on the rotated surface code the decoder outperforms correlated MWPM under both phenomenological and circuit-level noise, with up to $25\%$ fewer logical failures.
Secondly, on a family of high-rate qLDPC codes it matches BP+OSD on the smaller codes and surpasses it on the larger ones, with up to $16\%$ fewer logical failures on $[\![130,4,6]\!]$.
Thirdly, its decoding time does not depend on the physical error rate, whereas that of BP+OSD grows with it: on $[\![130,4,6]\!]$ its effective cost per shot is of the order of a few ms on a single GPU against a few tens of milliseconds for BP+OSD on a single CPU core, and its fixed computational graph makes it a candidate for real-time decoding on neutral-atom processors.
Finally, a single model trained across codes of different families outperforms uncorrelated MWPM on the graph-like codes seen during training and stays within $10\%$ of BP+OSD on the others, while it tends to fail to generalize to unseen codes, particularly larger ones.
Its probabilistic output enables confidence-based post-selection, lowering the logical error rate by more than an order of magnitude on the surface code while keeping more than $90\%$ of the shots. Since only the DEM changes, new codes and noise models can be decoded without redesigning the decoder.

\end{abstract}

\maketitle



\begin{figure*}[t]
    \centering
    \begin{subfigure}[t]{0.33\textwidth}
        \centering
        \includegraphics[width=\linewidth]{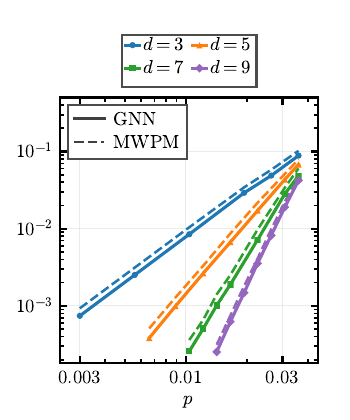}
        \caption{Rotated surface code, phenomenological noise, $d\in\{3,5,7,9\}$.}
        \label{fig:hero_sc}
    \end{subfigure}%
    \hfill%
    \begin{subfigure}[t]{0.33\textwidth}
        \centering
        \includegraphics[width=\linewidth]{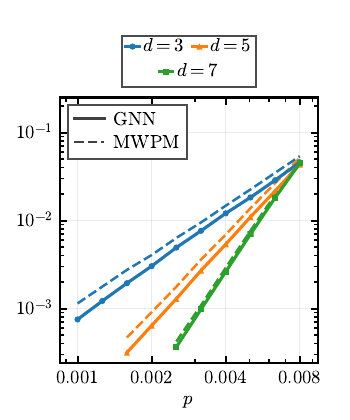}
        \caption{Rotated surface code, circuit-level noise, $d\in\{3,5,7\}$.}
        \label{fig:hero_circuit}
    \end{subfigure}%
    \hfill%
    \begin{subfigure}[t]{0.33\textwidth}
        \centering
        \includegraphics[width=\linewidth]{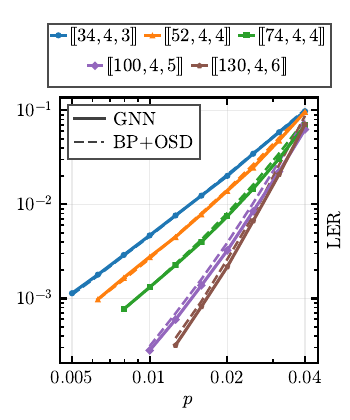}
        \caption{qLDPC LaCross $k{=}2$ codes, phenomenological noise, from
        $[\![34,4,3]\!]$ to $[\![130,4,6]\!]$.}
        \label{fig:hero_lacross}
    \end{subfigure}
    \caption{Logical error rate of the GNN decoder (solid) against the
    classical decoder (dashed) as a function of the physical error rate $p$, on identical
    shots and with $r{=}d$ syndrome rounds. The baseline is correlated minimum-weight perfect matching on the two surface-code panels and BP+OSD on the qLDPC panel.}
    \label{fig:hero_ler}
\end{figure*}

\section{Introduction}\label{sec:intro}

Quantum error correction (QEC) is a necessary ingredient of fault-tolerant quantum
computing~\cite{gottesmanIntroductionQuantumError2010,nielsenQuantumComputationQuantum2010,knillResilientQuantumComputation1998}. It exploits redundancy by encoding a small number of logical qubits in several noisy qubits, this procedure being feasible within a noise threshold that is set both by the decoder and the quantum code~\cite{shorSchemeReducingDecoherence1995,gottesman_stabilizer_1997,knillTheoryQuantumErrorcorrecting1997,preskillFaulttolerantQuantumComputation1998,steaneEfficientFaulttolerantQuantum1999}.
Among the many existing codes, the (rotated) surface code~\cite{kitaevQuantumComputationsAlgorithms1997,
fowlerSurfaceCodesPractical2012,dennisTopologicalQuantumMemory2002} stands out for its high threshold and short-range connectivity, while high-rate quantum low-density parity-check (qLDPC)
codes~\cite{pecorariHighRateQuantumLDPC2024} promise a much smaller qubit overhead on platforms with long-range connectivity. Crucial to QEC is the \emph{decoder}, devised to infer from syndrome measurements the recovery to be applied,
ideally in real time. Different types of decoders are indicated for different classes of codes. Minimum-weight
perfect matching (MWPM), implemented in
PyMatching~\cite{higgottPyMatchingPythonPackage2022,higgottSparseBlossomCorrecting2025}, is for example indicated for graph-like error models, where every fault flips at most two detectors. Such hurdle is partially tackled by its correlated variant~\cite{fowlerOptimalComplexityCorrection2013}.  
As hardware moves towards higher-rate codes and more realistic noise~\cite{bravyiHighthresholdLowoverheadFaulttolerant2024,googleQuantumErrorCorrection2025}, research in the decoding field becomes more and more active and new specialized classical decoders are devised~\cite{mullerImprovedBeliefPropagation2025,aghababaiebeniTesseractSearchbasedDecoder2025,wuMinimumweightParityFactor2025}; for example, belief propagation followed by ordered-statistics decoding (BP+OSD)~\cite{roffeDecodingAcrossQuantum2020,panteleevDegenerateQuantumLDPC2021} deals with qLDPC
codes, with a cost that grows with the code size and with the noise.
However, in this picture a decoder adaptable to arbitrary codes without redesign also becomes a priority. In this sense, machine learned decoders represent a viable option as they can in principle learn decoding strategies directly from data and potentially adapt to different code structures and noise models without requiring a different decoding algorithm for each code.

Training a decoder on QEC data is a task that is being dealt with for almost a decade now~\cite{torlaiNeuralDecoderTopological2017,krastanovDeepNeuralNetwork2017,varsamopoulosDecodingSmallSurface2018,chamberlandDeepNeuralDecoders2018,baireutherMachinelearningassistedCorrectionCorrelated2018,baireutherNeuralNetworkDecoder2019,niNeuralNetworkDecoders2020,andreassonQuantumErrorCorrection2019,swekeReinforcementLearningDecoders2021,meinerzScalableNeuralDecoder2022,gicevScalableFastArtificial2023};
see Ref.~\cite{leeMachineLearningApproaches2026} for a recent review. On the surface code the state
of the art is set by recurrent transformers, which outperform MWPM on experimental
data~\cite{bauschLearningHighaccuracyError2024} and decode in real time up to distance
$11$~\cite{seniorScalableRealtimeNeural2025}; graph neural networks (GNNs) on the detector graph
also outperform MWPM under circuit-level noise~\cite{langeDatadrivenDecodingQuantum2025} and
have been mapped to Field Programmable Gate Arrays (FPGA)~\cite{ciceroLowLatencyGNN2026}, and a transformer trained under
phenomenological noise approaches the optimal decoder at $d\in\{3,5\}$~\cite{seipMachineLearningOptimal2026}.
A systematic comparison of neural paradigms finds the inductive bias --- i.e. the set of assumptions a learning algorithm uses to predict outputs for inputs it has not encountered --- to be essential for accuracy rather than the architecture itself~\cite{yanRethinkRoleNeural2026}. 
For qLDPC codes, machine learning was first used to repair belief propagation
itself~\cite{liuNeuralBeliefPropagationDecoders2019,gongGraphNeuralNetworks2024}, and then to replace it with
graph~\cite{ninkovicDecodingQuantumLDPC2024}, hypergraph~\cite{bhaveHyperNQHypergraphNeural2025} and
Bayesian attention~\cite{miUncertaintyAwareGeneralizableNeural2025} networks on the Tanner graph.
Astra~\cite{maanMachineLearningMessagepassing2025} has been shown to outperform BP+OSD under code-capacity noise. At circuit level noise, a recurrent
transformer~\cite{blueMachineLearningDecoding2026} --- the code-agnostic
GraphQEC~\cite{huEfficientUniversalNeuralNetwork2025} --- and a convolutional decoder exploiting the
translation symmetry of the lattice~\cite{guScalableNeuralDecoders2026} were shown to surpass BP+OSD on
bivariate-bicycle codes (for Ref.~\cite{blueMachineLearningDecoding2026}, on the $[\![72,12,6]\!]$ code only). Generative~\cite{caoGenerativeDecodingQuantum2025,liuDecodingQuantumLow2025}
and ``foundation''~\cite{yanEfficientFoundationDecoders2026} decoders complete the picture. Neural
decoders further provide soft outputs~\cite{haugMachinelearnedSyndromePostselection2026} that are used to post-select~\cite{dentelskiNeuralNetworkDecoder2026},
and furthermore a cost independent of the syndrome that is attractive for real-time
decoding~\cite{battistelRealtimeDecodingFaulttolerant2023}. Table~\ref{tab:related} summarises the
works closest to ours. A common trait that emerges in most of these works is that decoders are tied to the specific code or a specific hardware. It would be instead highly desirable to have a decoder whose input is only the information an optimal decoder actually uses --- the syndrome, together with the structure of the noise and the prior probability of each fault --- and nothing specific to the code.

In this work we present POLYMECHANON, a GNN decoder whose only input is the detection error model (DEM) of a quantum
code under a given noise model~\cite{gidneyStimFastStabilizer2021}, represented as a tripartite
graph of detectors, error mechanisms and logical observables. Existing graph decoders place nodes on qubits and
checks~\cite{maanMachineLearningMessagepassing2025,huEfficientUniversalNeuralNetwork2025,ninkovicDecodingQuantumLDPC2024}
or on detectors alone~\cite{langeDatadrivenDecodingQuantum2025}, while convolutional and recurrent
decoders rely on a lattice~\cite{bauschLearningHighaccuracyError2024,guScalableNeuralDecoders2026}. Therefore, they do not fully exploit the fault structure contained in the DEM and in most cases they deploy code-dependent architectures. 
Instead, we build the graph from the DEM itself: a node is assigned to every detector, every error mechanism and every logical observable, and edges connect each error mechanism to the detectors it triggers and to the logical observables it flips. The error-mechanism nodes further carry their prior probability, i.e. the probability that the fault occurs before the syndrome is observed. Together with the measured syndrome, this is all the information an optimal decoder needs: the noise model becomes part of the input rather than of the architecture, and faults that trigger more than two detectors are represented as they are, without decomposition.

We benchmark the decoder on the rotated surface code under phenomenological and circuit-level
noise and on high-rate LaCross codes under phenomenological noise, against correlated MWPM and BP+OSD. The probabilistic output enables confidence-based post-selection, and remarkably its
decoding cost does not depend on the physical error rate. 

There are two main advances in this work. Firstly, we show that the detection error model alone is sufficient input for a neural decoder: with a fixed recipe for the graph and its features, and without any code-specific inductive bias --- no lattice, no translation symmetry, no assumption on the stabiliser weight --- the same architecture reaches or exceeds the strongest classical baselines on two structurally different code families and two noise models, namely correlated MWPM on the surface code and BP+OSD on high-rate qLDPC codes. Moreover, within the LaCross family the advantage over BP+OSD grows with the size of the code, from parity on $[\![34,4,3]\!]$ to $16\%$ fewer logical failures on $[\![130,4,6]\!]$ (Fig.~\ref{fig:lambda_size}). Secondly, we tackle the latency of decoding high-rate qLDPC codes: the cost of our decoder does not grow with the noise strength, and it remains up to one order of magnitude cheaper than BP+OSD, which makes it a promising building block towards real-time decoding. Finally, our work leaves open the theme of generalisation across graphs.
A single model trained on fifteen codes of different families decodes the codes seen in training close to or better than their classical baselines, and transfers to unseen codes whose size is comparable; its accuracy, however, drops as the target graph grows beyond the largest one seen during the training, a limit that neither more codes nor more parameters removed in our tests.
Since this limit is set by the size of the graph rather than by the code family, it suggests that the network relies on the fault structure of the DEM rather than on family-specific features. 
Given the partial nature of these tests, however, generalisation across graphs remains an open question in our work.

The paper is organised as follows:
Sec.~\ref{sec:background} reviews the codes, the noise models and the DEM;
Sec.~\ref{sec:decoder} describes the GNN architecture and its training; results are then presented in Sec.~\ref{sec:results}; Sec.~\ref{sec:outlook} discusses limitations and outlook. Finally, Appendix~\ref{sec:appendix_stim} details the DEM construction and Appendix~\ref{sec:appendix_arch} is devoted to specifics on the GNN architecture.

\begin{figure*}[t]
    \centering

    \begin{subfigure}[t]{0.48\textwidth}
        \centering
        \includegraphics[width=0.75\linewidth]{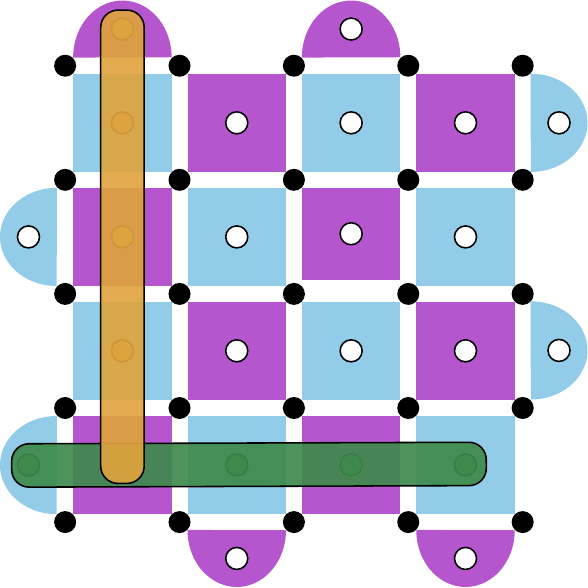}
        \caption{Rotated surface code of distance $d=5$. Every {\color{errorcol} $X$}- and {\color{detector} $Z$}-stabilizer is displayed, along with the only pair of logical operators {\color[HTML]{3E894C}$X_L$} and {\color[HTML]{dfa33d}$Z_L$}.}
        \label{fig:surfacecode}
    \end{subfigure}
    \hfill
    \begin{subfigure}[t]{0.50\textwidth}
        \centering
        \includegraphics[width=0.75\linewidth]{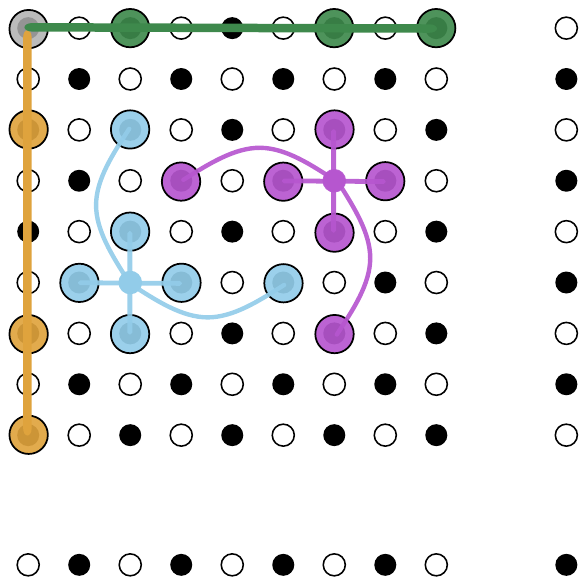}
        \caption{
        Example of LaCross code with next-to-nearest neighbor connectivity. An {\color{errorcol} $X$}- and a {\color{detector} $Z$}-stabilizer are depicted, along with one pair of logical {\color[HTML]{3E894C}$X_L$} and {\color[HTML]{dfa33d}$Z_L$} operators.
        }
        \label{fig:lacross}
    \end{subfigure}

    \caption{The two code families studied in this work.}
    \label{fig:codes}
\end{figure*}


\begin{table*}[t]
\caption{Neural decoders closest to this work. \emph{Graph / input}: what the network
operates on. \emph{Noise}: CC = code capacity, Ph =
phenomenological, CL = circuit level, Exp = experimental data. \emph{Training}: whether a
separate model is trained per code (and distance) or a single model is shared.}
\label{tab:related}
\begin{ruledtabular}
\begin{tabular}{lp{4.4cm}p{3.3cm}lp{3.1cm}}
Work & Graph / input & Codes & Noise & Training \\
\hline
AlphaQubit~\cite{bauschLearningHighaccuracyError2024,seniorScalableRealtimeNeural2025}
  & \rl recurrent transformer on the stabiliser lattice & \rl surface, colour & CL, Exp & \rl per code \\
Lange \emph{et al.}~\cite{langeDatadrivenDecodingQuantum2025}
  & \rl GNN on the detector graph & \rl surface & CL & \rl per code \\
Seip \emph{et al.}~\cite{seipMachineLearningOptimal2026}
  & \rl transformer on stabiliser measurements & \rl surface & CC, Ph, CL & \rl per code and noise value \\
Astra~\cite{maanMachineLearningMessagepassing2025}
  & \rl GNN on the Tanner graph & \rl surface, bivariate bicycle & CC & \rl per code, extrapolated in $d$ \\
Blue \emph{et al.}~\cite{blueMachineLearningDecoding2026}
  & \rl recurrent transformer & \rl bivariate bicycle & CL & \rl per code \\
GraphQEC~\cite{huEfficientUniversalNeuralNetwork2025}
  & \rl GNN on the Tanner graph with logical nodes & \rl surface, colour, bivariate bicycle & CL, Exp & \rl per code \\
Gu \emph{et al.}~\cite{guScalableNeuralDecoders2026}
  & \rl convolution on the code lattice & \rl surface, bivariate bicycle & CL & \rl per code \\
Yan \emph{et al.}~\cite{yanEfficientFoundationDecoders2026}
  & \rl transformer & \rl surface, bivariate bicycle & CL & \rl transferred across $d$ \\[4pt]
\textbf{POLYMECHANON}
  & \rl GNN on the DEM: detectors, error mechanisms, logicals & \rl surface, LaCross ($+$ other codes, exploratory) & Ph, CL & \rl per code; one model for different codes \\
\end{tabular}
\end{ruledtabular}
\end{table*}

\section{Background}\label{sec:background}

In this chapter we outline three ingredients necessary to quantify and improve decoding performance:  a code, which defines the encoded logical information and how it is protected; a noise model, which specifies how physical errors arise; and a representation of the resulting fault structure on which a decoder can act. We first define the decoding problem and the figures of merit used throughout the paper (Sec.~\ref{sec:figures_of_merit}), then introduce the two code families under study (Sec.~\ref{sec:codes_under_study}), and subsequently the noise models applied to them (Sec.~\ref{sec:noise_models}). Finally, we conclude with the detection error model and its tripartite-graph representation (Sec.~\ref{sec:dem}), which constitutes the input to the decoder introduced in Sec.~\ref{sec:decoder}.

\subsection{Figures of merit}\label{sec:figures_of_merit}

In order to discuss specific codes and the meaning of their decoding process, it is useful to outline the quantities that quantify decoding performance. Some technical terms appearing here will be defined and clarified in the following sections.

A stabiliser code of distance $d$ can correct errors acting on at most $\lfloor (d-1)/2 \rfloor$ qubits; equivalently, $d$ is the minimum weight of a \textit{logical operator}, i.e.\ of an undetectable error that alters the encoded information~\cite{gottesmanIntroductionQuantumError2010,nielsenQuantumComputationQuantum2010}. Its \textit{weight} is the number of qubits on which it acts in a non-trivial fashion.

The quality of a decoder is quantified by the \textit{logical error rate} (LER): the probability that the decoder's correction, combined with the actual error, results in a logical operation. For a code of distance $d$ operated at a noise strength $p$ below a \textit{threshold} $p_\mathrm{thr}$, the LER scales as~\cite{dennisTopologicalQuantumMemory2002,fowlerSurfaceCodesPractical2012,wangThresholdErrorRates2010}
\begin{equation}
    \mathrm{LER} \sim \left(\frac{p}{p_\mathrm{thr}}\right)^{(d+1)/2},
    \label{eq:ler_scaling}
\end{equation}
so that increasing $d$ suppresses the logical error exponentially; above $p_\mathrm{thr}$, increasing $d$ provides no longer an advantage. How the noise strength $p$ enters in the noise model is described in Appendix~\ref{sec:appendix_stim}. Both the threshold and the pre-factor of a given decoder's LER curve are decoder-dependent, and a better decoder improves them --- but only up to intrinsic limits set by the code. The optimal threshold is a property of the code and noise model, not of the decoder. As an example to this statement, it was recently shown in Ref.~\cite{english_duality_2026} that the optimal threshold under code-capacity noise for (zero-rate) \textit{em}-symmetric CSS codes --- i.e. such that their two parity-check matrices are related through column and row permutations --- is constrained, to a single universal value.

In this work, we focus on so-called \textit{memory-$Z$
experiments}~\cite{fowlerSurfaceCodesPractical2012,googleQuantumErrorCorrection2025}, where the logical qubit(s) is (are) initialised in the $+1$ eigenstate(s) $\ket{0_L}$ of the logical operator(s) $Z_L$, $r=d$ rounds of syndrome extraction are performed, and the decoder must determine whether a logical error $X_L$ has occurred. Initialising in the $Z_L$ basis makes the logical qubit insensitive to $Z$-errors, and hence to $Z_L$ logical errors, while it still must be protected against $X$ (and $X_L$) errors. The symmetric case is realised by initialising in the $+1$ eigenstate(s) $\ket{+_L}$ of $X_L$.

The concepts expounded in this paragraph apply to any stabiliser code. We now introduce the two
families on which we perform the decoding task.

\subsection{Codes under study}\label{sec:codes_under_study}

We study two distinct code families (Fig.~\ref{fig:codes}): a topological code with local stabilizers --- the rotated surface code --- and a high-rate  qLDPC family with non-local stabilizers --- the LaCross codes. Both are Calderbank--Shor--Steane (CSS) codes, and sharing a single decoder across both is the crucial challenge addressed in this work. More generally, our decoder applies to any stabilizer code that can be compiled into a detection error model (Section~\ref{sec:dem}).

In Section~\ref{sec:transfer} we comment on a ``generalist'' model trained on a variety of different families of codes that however we do not define in this paragraph. 

\textit{CSS Codes.}
CSS codes are stabiliser
codes~\cite{gottesman_stabilizer_1997,gottesmanIntroductionQuantumError2010}
whose generators are tensor products of either $X$-type or $Z$-type Pauli
operators.
Their parity-check matrix $H_Z \in \mathbb{F}_2^{M\times 2N}$ is the antidiagonal symplectic matrix
\begin{equation}
    H = \begin{pmatrix} 0 & H_Z \\ H_X & 0 \end{pmatrix},
    \qquad H_X H_Z^{\top} = 0 ,
    \label{eq:css}
\end{equation}
where the matrix $H_X\in \mathbb{F}_2^{M\times N}$ ($H_Z\in \mathbb{F}_2^{M\times N}$) acts on the $N$ physical qubits via the action of the $M$ stabilizers $X_1,\dots,X_M$($Z_1,\dots,Z_M$). It is not required that the stabilizers be linearly independent. Both $H_X$ and $H_Z$ are respectively the parity-check matrix of a classical linear
code~\cite{macwilliams_theory_1977}, and the CSS construction combines two such classical codes into a quantum one, the commutation condition $H_X H_Z^{\top}=0$ ensuring that all stabilizers commute~\cite{calderbank_good_1996,steane_error_1996}. 
This is where classical coding theory enters quantum error correction: bit-flip ($X$) and phase-flip ($Z$) errors are detected exactly as in the classical codes defined by $H_Z$ and $H_X$, so that decoding a CSS code reduces to two classical decoding problems, syndrome extraction decomposes into two separate circuits and logical gates admit transversal implementations.
The two families studied here are \emph{hypergraph
products}~\cite{tillich_quantum_2014} of simple classical codes --- a
repetition code for the surface code and a weight-three cyclic code for LaCross.

\textit{Rotated surface code.}
The rotated surface code~\cite{kitaevQuantumComputationsAlgorithms1997,fowlerSurfaceCodesPractical2012,horsmanSurfaceCodeQuantum2012,bombinOptimalResourcesTopological2007} is the CSS hypergraph product of two copies of a classical repetition code
$[n,1,n]$, yielding weight-four stabilizer checks that act only on neighbouring
data qubits. A distance-$d$ code
encodes one logical qubit in $d^2$ data qubits, with a further overhead of $d^2-1$ qubits serving as ancillas for the measurement of the $X-$ and $Z-$type stabilizers (cfr. Fig.~\ref{fig:surfacecode}), and has parameters $[\![d^2,1,d]\!]$. Its logical operators $X_L$ and $Z_L$ are Pauli strings connecting opposite boundaries of the lattice. Due to the local connectivity and high threshold, the surface code is the standard model considered in QEC.

\textit{LaCross codes.}
LaCross codes are qLDPC codes ~\cite{roffeDecodingAcrossQuantum2020,pecorariHighRateQuantumLDPC2024} obtained via the same
hypergraph-product construction, but replacing the repetition code with an equal cyclic classical code with generating polynomial $h(x)=~1 +~ x +~x^k\mathds{1}_{\{k>1\}}$~\cite{pecorariHighRateQuantumLDPC2024}.
For $k\ge 2$ this promotes the weight-two checks of the repetition code to
weight-three checks, giving weight-six stabilizers and code parameters
$[\![n^2+(n-k)^2,k^2,d]\!]$ with open boundary conditions. 
For $k=1$ the construction yields codes that are equivalent to the (unrotated) surface code, thus encoding a single logical qubit; an example with $k=2$ is drawn in Fig.~\ref{fig:lacross}. For $k\ge 2$ the codes become genuinely high-rate: $k=2$ gives $k^2=4$ logical qubits and $k=3$ gives $k^2=9$. These are the primary LaCross targets, of particular interest for neutral-atom platforms with long-range connectivity~\cite{pecorariAddressableGatebasedLogical2025, pecorariQuantumLowdensityParitycheck2025,bergonzoni_fault-tolerant_2026}.

The two families of codes having been introduced, we describe next the type of noise injected in the simulations.

\subsection{Noise models}\label{sec:noise_models}

\begin{figure}[t]
\centering
\includegraphics[width=\linewidth]{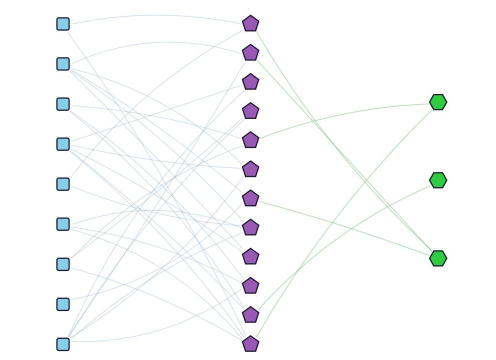}
\caption{Representation of an extended Tanner graph. From left to
    right, the nodes are the {\color{detector}detectors} ($\nD$), the
    {\color{errorcol}error mechanisms} ($\nEM$), and the {\color{logical}logical
    observable} ($\nL$) of the detection error model of a surface code. Edges encode
    which detectors are triggered by each error mechanism and which mechanisms
    flip the logical observable. An even number of triggers has no effect in the noise regimes that we operate with.}
\label{fig:toy_tripartite_graph}
\end{figure}

\begin{table}[t]
\caption{Tripartite graph sizes with $r{=}d$ syndrome rounds, as seen by the
    decoder. $N_{\nD}$: detectors, $N_{\nEM}$: error mechanisms, $N_{\nL}$: logical observables,
    $|E|$: total edges ($|E_{\nD,\nEM}|+|E_{\nEM,\nL}|$); $f_{>2}$: fraction of error mechanisms
    that trigger more than two detectors, i.e.\ hyperedges.}
\label{tab:graph_size}
\begin{ruledtabular}
\begin{tabular}{lrrrrrr}
    Code & $N_{\nD}$ & $N_{\nEM}$ & $N_{\nL}$ & $|E|$ & Nodes & $f_{>2}$ \\
\hline
    \multicolumn{7}{l}{\textit{Rotated surface code, phenomenological}} \\
    $[\![9,1,3]\!]$    & 32    & 107     & 1 & 219      & 140     & 14\% \\
    $[\![25,1,5]\!]$   & 144   & 509     & 1 & 1\,130   & 654     & 21\% \\
    $[\![49,1,7]\!]$   & 384   & 1\,399  & 1 & 3\,213   & 1\,784  & 23\% \\
    $[\![81,1,9]\!]$   & 800   & 2\,969  & 1 & 6\,948   & 3\,770  & 23\% \\[4pt]
    \multicolumn{7}{l}{\textit{Rotated surface code, circuit level}} \\
    $[\![9,1,3]\!]$    & 32    & 333     & 1 & 934      & 366     & 54\% \\
    $[\![25,1,5]\!]$   & 144   & 2\,087  & 1 & 6\,454   & 2\,232  & 66\% \\
    $[\![49,1,7]\!]$   & 384   & 6\,361  & 1 & 20\,390  & 6\,746  & 70\% \\
    $[\![81,1,9]\!]$   & 800   & 14\,259 & 1 & 46\,582  & 15\,060 & 71\% \\[4pt]
    \multicolumn{7}{l}{\textit{LaCross $k{=}2$, phenomenological}} \\
    $[\![34,4,3]\!]$   & 120   & 386     & 4 & 1\,091   & 510     & 45\% \\
    $[\![52,4,4]\!]$   & 240   & 804     & 4 & 2\,396   & 1\,048  & 52\% \\
    $[\![74,4,4]\!]$   & 350   & 1\,154  & 4 & 3\,535   & 1\,508  & 56\% \\
    $[\![100,4,5]\!]$  & 576   & 1\,964  & 4 & 6\,136   & 2\,544  & 59\% \\
    $[\![130,4,6]\!]$  & 882   & 3\,078  & 4 & 9\,783   & 3\,964  & 61\% \\[4pt]
    \multicolumn{7}{l}{\textit{LaCross $k{=}3$, phenomenological}} \\
    $[\![65,9,4]\!]$   & 280   & 983     & 9 & 2\,799   & 1\,272  & 42\% \\
    \end{tabular}
\end{ruledtabular}
\end{table}

Decoder performance is meaningful only when specified in respect to an assumed error model. We now expound the standard hierarchy of noise
regimes in increasingly realistic order~\cite{stephensFaulttolerantThresholdsQuantum2014,ioliusDecodingAlgorithmsSurface2024}:
\textit{code capacity} (data-qubit depolarising noise only, perfect
measurements); \textit{phenomenological} (data-qubit depolarising noise together with measurement
bit-flip noise with equal probability $p$, over $r=d$ rounds); and \textit{circuit
level} (noise with probability $p$ after every gate, measurement and reset operation, over $r=d$ rounds). For more comments, we refer the reader to Appendix~\ref{sec:appendix_stim}.

This work covers the phenomenological model for both code families and, for the surface code, the circuit-level model as well. The difference between the two is not merely quantitative: at circuit level a single fault propagates through the entangling gates and can light up more than two detectors, so that a large fraction of the error mechanisms
falls outside the graphlike class that MWPM decoders are able to represent~\cite{higgottPyMatchingPythonPackage2022,higgottSparseBlossomCorrecting2025}.

Under the phenomenological model used here, we measure a surface-code threshold $p_\mathrm{thr}^{\mathrm{SC}} \approx 0.041$ under
MWPM, consistent with the classic $\approx\!0.029$ independent-bit-flip value~\cite{wangConfinementHiggsTransition2003,stephensFaulttolerantThresholdsQuantum2014} within the difference in noise convention.
The LaCross $k{=}2$ family reaches a comparable threshold $p_\mathrm{thr}^{\mathrm{LC}} \approx 0.037$ under
BP+OSD, obtained from the crossing of the logical-error-rate curves for $d\in\{3,4,5\}$.
Under circuit-level noise the surface-code threshold drops to
$p_\mathrm{thr}^{\mathrm{SC,cl}} \approx 0.0075$.

Given a code and a noise model, the fault structure of the entire experiment is
compiled into a compact, code-agnostic object --- the detection error model --- which is the input of the decoder.

\subsection{Detection error model and tripartite graph}\label{sec:dem}

Having fixed the code and the noise model, we outline their
representation on which the decoder acts. Any stabiliser code can be compiled into a circuit from which \texttt{stim}~\cite{gidneyStimFastStabilizer2021} extracts a \emph{detection error model} (DEM) --- a compressed description of all the syndromes and fault paths. The decoder operates solely on the DEM (see Appendix~\ref{sec:appendix_stim} for circuit construction details). The DEM consists of:
\begin{itemize}
    \item \textit{detectors}, defined as XOR of consecutive stabilizer measurements (cfr. App.~\ref{sec:appendix_stim}), and numbered $1,\dots,N_{\nD}$;
    \item \textit{error mechanisms}, each with an assigned prior probability, and numbered $1,\dots,N_{\nEM}$;
    \item \textit{logical observables}, and numbered $1,\dots,N_{\nL}$;
    \item a binary matrix $H \in \mathbb{F}_2^{N_{\nD} \times N_{\nEM}}$, where $H_{ij}=1$
          if the $j-$th error mechanism flips the $i-$th detector;
    \item a binary matrix $L \in \mathbb{F}_2^{N_{\nL} \times N_{\nEM}}$, where $L_{ij}=1$ if the $j-$th error mechanism flips the $i-$th logical observable. For the surface code $L$ reduces to a binary vector in $\mathbb{F}_2^{1\times N_{\nEM}}$, as a single logical observable is encoded.
\end{itemize}

A \emph{syndrome} is the binary vector $\sigma$ listing all the detectors that have fired during one round of error correction, i.e. a single cycle of stabilizer extraction. We indicate with $\sigma$ the complete history $\sigma=(\sigma_1,\dots,\sigma_r)\in \mathbb{F}_2^{N_{\nD}}$ of syndromes throughout full cycles of error correction, composed of $r$ rounds of single cycles. 
Similarly, for notation convenience we represent error mechanisms of a full cycle of error correction by means of vectors $e\in\mathbb{F}_2^{N_{\nEM}}$, where $e_j=1$ if and only if the $j$-th error mechanism occurred. In an analogue way, a logic observable is a vector $\ell\in\mathbb{F}_2^{N_{\nL}}$ accounting for the presence of logical errors. In the introduced formalism, an error $e$ generates the syndrome $\sigma = H\cdot e \pmod{2}$, and decoding amounts to inferring whether $\ell = L\cdot e \pmod 2$ from $\sigma$. In the following, whenever making computations between vectors and matrices with only $0$s and $1$s as entries, the term $\pmod{2}$ will be omitted but implied.

The DEM treats error mechanisms as independent between each others, so that the prior probability over error configurations $e=(e_1,\dots,e_{N_{\nEM}})\in\mathbb{F}_2^{N_{\nEM}}$ factorises as
\begin{equation}
    P(e) = \prod_{j=1}^{N_{\nEM}} p_j^{\,e_j}\,(1-p_j)^{\,1-e_j},
    \label{eq:prior}
\end{equation}
where $p_j$ is the prior probability of the single $j$-th mechanism. Each mechanism is equivalently described by the logit of its prior probability, \begin{equation}\label{eq:loglike}\lambda_j = \log\!\big(p_j/(1-p_j)\big),\end{equation} the quantity carried by the corresponding $\nEM$-node, which we define later in this paragraph. Given a syndrome $\sigma$, a maximum-likelihood decoder returns the most probable of two scenarios, that the error $e$ triggered a logical error or that it did not. This conditioned probability is computed as
$\sigma$~\cite{dennisTopologicalQuantumMemory2002,ioliusDecodingAlgorithmsSurface2024}
\begin{equation}
    P(\ell \mid \sigma) = \frac{1}{P(\sigma)}
    \sum_{\substack{e \in \mathbb{F}_2^{N_{\nEM}}: \\ H \cdot e = \sigma}}
    \mathds{1}_{\{L\cdot e = \ell\}}(e)\; P(e),
    \label{eq:marginal}
\end{equation}
where $P(\sigma)=\sum_{e:\,H\cdot e=\sigma} P(e)$ and $\mathds{1}_{\{L \cdot e = \ell\}}(e)$ is $1$ if $L \cdot e = \ell$ and $0$ if not. 
In the sum, the index $e$ is an error mechanism whose syndrome $H \cdot e$ equals the input syndrome $\sigma$, so that every error $\tilde e$ with the property $H\cdot \tilde e\not=\sigma$ is not summed over.
The decoded outcome should be $\hat{\ell} = \arg\max_{\ell} P(\ell \mid \sigma)$, i.e. the logical operator $\hat\ell$ that for a given syndrome $\sigma$ is most likely to have happened. However, the sum runs over an exponentially large number of error configurations, making the
exact maximization \textit{intractable} in general. This unfeasible regime gives the motivation to consider approximate decoders such as belief propagation~\cite{poulinIterativeDecodingSparse2008,oldGeneralizedBeliefPropagation2023} and machine-learned decoders as the one presented in this work, where a graph
neural network learns to approximate Eq.~\eqref{eq:marginal} directly on its associated graph.

The DEM has a natural tripartite structure that we make explicit by representing it as a tripartite graph $\mathcal{G}=(\mathcal{V},\mathcal{E})$ --- $\mathcal{V}$ and $\mathcal{E}$ being the vertices and the edges of $\mathcal{G}$ respectively. Specifically, $\mathcal{V}=\nD\cup\nEM\cup \nL$ and $\mathcal{E}=E_{\nD,\nEM}\cup E_{\nEM,\nL})$ (Fig.~\ref{fig:toy_tripartite_graph}) as described in what follows:
\begin{itemize}
    \item[$\star$] \textit{$\nD$-nodes} (detectors): one per detector, carrying the syndrome bit $0/1$;
    \item[$\star$] \textit{$\nEM$-nodes} (error mechanisms): one per fault mechanism. These nodes are initialized with the logit $\lambda_j$ of their prior probability;
    \item[$\star$] \textit{$\nL$-nodes} (logical observables): one per logical qubit. As for the error mechanisms, they are initially endowed with the logits of their prior probabilities.
\end{itemize}
Finally, edges $E_{\nD,\nEM}$ and $E_{\nEM,\nL}$ encode respectively the matrices $H$ and $L$ previously defined. This representation makes the belief-propagation structure~\cite{oldGeneralizedBeliefPropagation2023} of the problem
explicit: decoding consists in determining the probabilities of $\nEM$-nodes and $\nL$-nodes in $\mathcal{G}$, having as inputs solely the syndrome bits for the $\nD$-nodes and the prior probabilities in the $\nEM$-nodes and in the $\nEM$-nodes and the $\nL$-nodes. That is, the probabilistic information depending only on the noise model and on the specific faults.

Table~\ref{tab:graph_size} reports the resulting graph sizes across codes and noise models considered in this work; the number of error-mechanism nodes and edges grows rapidly with the code, and dominates the decoder's cost.

So far we have illustrated the data that our decoder digests. In the following section we introduce our decoder by outlining how it elaborates such data, its architecture and the training process. 


\section{GNN Decoder}\label{sec:decoder}

\begin{table}[t]
\caption{Nine-dimensional node features, computed from the DEM alone --- the
matrices $H$ and $L$, the prior probabilities and the detector coordinates, the latter normalised
to their bounding box. Notation: $\mathcal{N}(\cdot)$ is the set of neighbours of a node in $\mathcal{G}$, so that $\mathcal{N}(d)$ are the mechanisms that flip a detector $d$, $\mathcal{N}(e)$ the detectors and logical observables flipped by the mechanism $e$,
$\mathcal{N}(l)$ the mechanisms in the support of the logical observable $l$; $\mathcal{N}^2(d)$ and $\mathcal{N}^2(l)$ are the sets of logical observables reachable by a two-edge path from a detector $d$ and vice versa, respectively; $\bar\sigma_S$ is the mean of the syndrome bits over a set $S$ of
detectors and $\lambda_e=\mathrm{logit}\,p_e$ the logit of the prior probability of Eq.~\eqref{eq:loglike}. In the following we describe the features per-node.
$\nD$-nodes: syndrome bit; stabiliser type ($X$ or $Z$), read from the parity of the detector
coordinates; temporal role (first, middle or last round); number of incident mechanisms, divided by
$4$; position relative to the bounding box and normalised time; syndrome averaged over the set $S$ of
detectors  that share a mechanism with $d$, each weighted by the number of shared mechanisms; the indicator function of $|\mathcal{N}(d)|<4$, which marks a boundary stabiliser. $\nEM$-nodes: logit of the prior probability;
measurement-error flag; number of detectors it flips, divided by $4$; the averaged coordinates of those detectors in
space and time, normalized; fraction of them that were triggered during the full cycle; indicator function that outputs $1$ if the mechanism $e$ is connected to a logical observable and $0$ if not; $\lambda_j$ relative to its median over all mechanisms. $\nL$-nodes: logit relative to its prior probability; number of neighboring $\nEM$-nodes, relative to the size of $N_{\nEM}$;
mean and maximum logits over all its neighboring $\nEM$-nodes; average and parity of the whole syndrome (clearly these numbers are the same for every $\nL$-node; their function is mostly to pad the dimension of logical observable so to have the same number as for the other nodes); weight $w(l)$ of the observable $l$, normalised; syndrome density over the detectors it reaches; and the size of that
reachable set, relative to $N_{\nD}$.}
\label{tab:features}
\begin{ruledtabular}
\begin{tabular}{clll}
    Dim & $\nD$-node $d$  & $\nEM$-node $e$  & $\nL$-node $l$ \\
\hline
    0 & $\sigma_i \in \{0,1\}$                & $\lambda_e$                             & $\lambda_l$  \\
    1 & coord.\ parity                        & meas.-error flag                         & $|\mathcal{N}(l)|/N_{\nEM}$ \\
    2 & temporal role                         & $|\mathcal{N}(e)\cap\nD|/4$                  & $\mathrm{mean}_{e\in\mathcal{N}(l)}\,\lambda_e$ \\
    3 & $|\mathcal{N}(d)|/4$                  & $x$                                              & $\max_{e\in\mathcal{N}(l)}\,\lambda_e$ \\
    4 & $x$                                   & $y$                                           & $\bar\sigma_{\nD}$ \\
    5 & $y$                                   & $t$                                          & $\bigoplus_i \sigma_i$ \\
    6 & $t$                                   & $\bar\sigma_{\mathcal{N}(e)\cap\nD}$                 & $w(l)/N_\nL$ \\
    7 & $\bar\sigma_S$                        & $\mathds{1}_{\{\mathcal{N}(e)\cap\nL\not=\emptyset\}}$      & $\bar\sigma_{\mathcal{N}^2(l)}$ \\
    8 & $\mathds{1}_{\{|\mathcal{N}(d)|<4\}}$ & $\lambda_e - \mathrm{median}(\nEM))$             & $|\mathcal{N}^2(l)|/N_{\nD}$ \\
    \end{tabular}
\end{ruledtabular}
\end{table}

\begin{table*}[t]
\caption{Per-code training hyperparameters: hidden dimension $F$,
message-passing iterations $T$, learning rate, and training schedule --- optimisation steps for
the streaming pipeline of Sec.~\ref{sec:training}, epochs for the models
inherited from the earlier one (per curriculum stage for the surface code, in a single stage over a fixed mixture of noise rates for LaCross).}
\label{tab:hyperparams}
\begin{ruledtabular}
\begin{tabular}{llcccl}
    Family & Code & $F$ & $T$ & LR & Schedule \\
\hline
    \multirow{4}{*}{Rot.\ SC (phenom.)}
    & $d{=}3$ & 64  & 6  & $5{\times}10^{-4}$, $5{\times}10^{-5}$ & 100 ep./stage${}+{}$100k \\
    & $d{=}5$ & 80  & 10 & $3{\times}10^{-4}$, $5{\times}10^{-5}$ & 100 ep./stage${}+{}$150k \\
    & $d{=}7$ & 128 & 10 & $3{\times}10^{-4}$, $10^{-4}$, $5{\times}10^{-5}$ & 100 ep./stage${}+{}$150k${}+{}$150k \\
    & $d{=}9$ & 200 & 14 & $5{\times}10^{-5}$, $10^{-4}$, $5{\times}10^{-5}$ & 80 ep./stage${}+{}$120k${}+{}$150k \\[4pt]
    \multirow{3}{*}{Rot.\ SC (circ.\ lev.)}
    & $d{=}3$ & 128 & 10 & $3{\times}10^{-4}$, $5{\times}10^{-5}$ & 60k${}+{}$100k steps \\
    & $d{=}5$ & 150 & 10 & $3{\times}10^{-4}$, $5{\times}10^{-5}$ & 60k${}+{}$100k steps \\
    & $d{=}7$ & 150 & 10 & $3{\times}10^{-4}$, $1.5{\times}10^{-4}$ & 60k${}+{}$150k${}+{}$300k steps \\[4pt]
    \multirow{5}{*}{LaCross $k{=}2$ (phenom.)}
    & $[\![34,4,3]\!]$  & 128 & 10 & $10^{-4}$ & 1500 ep. \\
    & $[\![52,4,4]\!]$  & 128 & 10 & $10^{-4}$ & 1500 ep. \\
    & $[\![74,4,4]\!]$  & 150 & 10 & $10^{-4}$ & 1500 ep. \\
    & $[\![100,4,5]\!]$ & 150 & 10 & $10^{-4}$ & 1500 ep. \\
    & $[\![130,4,6]\!]$ & 150 & 10 & $1.5{\times}10^{-4}$ & 60k steps \\[4pt]
    LaCross $k{=}3$ (phenom.) & $[\![65,9,4]\!]$  & 150 & 10 & $3{\times}10^{-4}$ & 60k steps \\[4pt]
    Multi-code (phenom.) & 15 codes & 150 & 10 & $3{\times}10^{-4}$ & 60k steps \\
    \end{tabular}
\end{ruledtabular}
\end{table*}

We now present our decoder: a graph neural network (GNN) that learns to
approximate Eq.~\eqref{eq:marginal} directly on the tripartite graph $\mathcal{G}$ of Sec.~\ref{sec:dem}, thereby bypassing the intractable exact sum over error configurations. 
Every data the network receives --- the syndrome bits, the prior probabilities of the error mechanisms, the edge structure encoded in $H$ and $L$, and a
small set of static features derived from them and from the detector coordinates that
\texttt{stim} attaches to the DEM (Sec.~\ref{sec:features}) --- is computed by one fixed recipe,
identical for every code and every noise model. Nothing in the architecture of the GNN  refers to a particular family, lattice, stabiliser
weight or number of logical qubits, and moving from the surface code to LaCross codes, or from
phenomenological to circuit-level noise, changes only the DEM and nothing else. The only per-code
choices are the number of hidden features $F$ and of iterations $T$, which set the capacity of the model.

We emphasize that, while we claim that any code admitting a DEM can, in principle, be decoded by training this network on the code without any architectural redesign, a separate question is whether a single set of weights can decode multiple code distances or even different families of quantum codes, regardless of whether those codes were included in the training set.
We test the first aspect on two code families and two noise
models (Secs.~\ref{sec:sc_results}--\ref{sec:lacross_results}), and the second with one network
shared across fifteen codes (Sec.~\ref{sec:transfer}).

The decoder proceeds in three stages. Each node of $\mathcal{G}$ is first equipped with a feature vector encoding its structural role within the graph and, for $\nD$-nodes, the observed syndrome (Sec.~\ref{sec:features}). These features are embedded and then refined over $T$ iterations of message passing with a recurrent update, letting information propagate between $\nD$-nodes, $\nEM$-nodes and $\nL$-nodes (Sec.~\ref{sec:architecture}). Finally, a logical-error probability is read out from the logical observables --- the $\nL$-nodes. A separate model is trained per code and distance, on a stream of newly sampled shots spanning the whole sub-threshold noise range (Sec.~\ref{sec:training}). In Appendix~\ref{sec:appendix_arch} we detail the single per-layer equations.

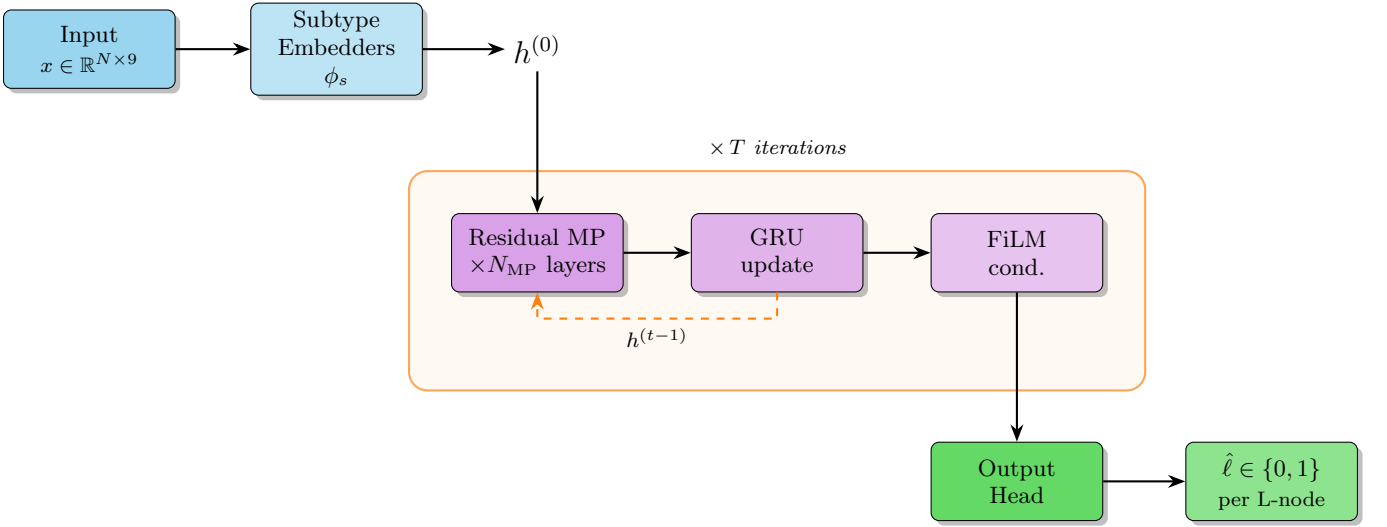
\begin{figure*}[!tp]
\centering
\resizebox{\linewidth}{!}{%
\begin{tikzpicture}[
    >=Stealth,
    box/.style={draw, rounded corners=3pt, minimum width=2.3cm,
                minimum height=1.05cm, align=center, font=\small,
                fill=white, drop shadow},
    sym/.style={font=\large},
    arr/.style={-Stealth, thick},
    feed/.style={-Stealth, thick, dashed, draw=gnnorange},
    iterbox/.style={draw=gnnorange!70, thick, rounded corners=7pt,
                    inner sep=16pt, fill=gnnorange!5},
    lbl/.style={font=\footnotesize\itshape, text=black}
]

\node[box, fill=detector!85] (inp) {Input\\[1pt]\footnotesize$x\in\mathbb{R}^{N\times 9}$};
\node[box, fill=detector!55, right=1.0cm of inp] (emb) {Subtype\\Embedders\\$\phi_s$};
\node[sym, right=1.1cm of emb] (h0) {$h^{(0)}$};
\draw[arr] (inp) -- (emb);
\draw[arr] (emb) -- (h0);

\node[box, fill=errorcol!55, below=1.9cm of h0]   (resmp) {Residual MP\\$\times \nMP$ layers};
\node[box, fill=errorcol!45, right=0.9cm of resmp] (gru)  {GRU\\update};
\node[box, fill=errorcol!35,  right=0.9cm of gru]    (film) {FiLM\\cond.};
\draw[arr] (resmp) -- (gru);
\draw[arr] (gru)   -- (film);

\begin{scope}[on background layer]
    \coordinate (fbbottom) at ($(gru.south)+(0,-0.75)$);   
    \node[iterbox, fit=(resmp)(film)(fbbottom)] (iter) {};
\end{scope}
\node[lbl, anchor=south] at ([yshift=2pt]iter.north) {$\times\,T$ iterations};

\draw[arr] (h0) -- (resmp);

\draw[feed] (gru.south) -- ++(0,-0.35) -| (resmp.south)
    node[pos=0.25, below, lbl] {$h^{(t-1)}$};

\node[box, fill=logical!75, below=2.0cm of film] (head) {Output\\Head};
\node[box, fill=logical!55, right=1.1cm of head]  (out)  {$\hat{\ell}\in\{0,1\}$\\[1pt]\footnotesize per $\nL$-node};
\draw[arr] (film.south) -- (head.north);
\draw[arr] (head) -- (out);

\end{tikzpicture}%
}

\caption{Overview of the POLYMECHANON forward pass. The nine-dimensional node
features are embedded via subtype-specific linear maps and non-linear activating functions. Over $T$
iterations, representations are refined by $\nMP$ residual message-passing layers and
updated by a GRU cell (Eq.~\ref{eq:gru_update}). After, a FiLM conditioning is applied. The final output
head maps the $\nL$-node representation to a logical error prediction.}

\label{fig:gnn_arch}
\end{figure*}

\subsection{Graph construction and features}\label{sec:features}

Every node in $\mathcal{G}$ is assigned a nine-dimensional feature vector (cfr. Table~\ref{tab:features} for more details). The features of a specific node in the graph are either static or dynamic, meaning that they might be the same or differ across different cycles of error correction. For instance, the temporal and spatial coordinates in a feature vector of a specific node do not vary through different cycles of error correction, since they are intrinsically related to the position in time and space of, e.g., a detector. Analogously, the feature coordinate encoding the prior probability of an error mechanism is static (if the noise strength is not varied) in the sense that, without any a-priori information, the probability of a specific error mechanism to happen is the same for any cycle of error correction. On the other hand, the feature coordinate of a node corresponding to a detector being $0$ (or $1$) obviously depends on the cycle of error correction.

Finally, spatial coordinates are normalised to the detector bounding box and temporal ones to
the interval $[0,1]$, so that the same features are defined, over the same range, for a code of
any size. All features are functions of the DEM --- its matrices $H$ and $L$, the prior probabilities and the
detector coordinates --- evaluated by the same recipe for every code. Some carry normalisations
chosen on the surface code (the number of neighbours of a node divided by $4$, the parity of the
detector coordinates); on other codes they act as generic connectivity and parity data, and a single network
trained across fifteen codes of different families uses them unchanged (Sec.~\ref{sec:transfer}).
Normalisation makes a trained network applicable to a code of a different size; whether it
is also accurate is an empirical question, which Sec.~\ref{sec:transfer} partially answers.

These vectors of features are the input to the network, whose architecture is described next.

\subsection{Architecture overview}\label{sec:architecture}

The network internal structure can be divided in three subparts: subtype-specific embedding, iterative message passing, and readout. Nodes are distinguished into six subtypes:
$\nD_\mathrm{init}$, $\nD_\mathrm{cycle}$, $\nD_\mathrm{final}$ (respectively first-, middle- and last-round detectors),
$\nEM_\mathrm{meas}$, $\nEM_\mathrm{data}$ (measurement- and data-error mechanisms), and $\nL$ (logical observables).
Each subtype has its own linear embedder --- i.e. a linear map $\phi_s: \mathbb{R}^9 \to \mathbb{R}^F$ ---
followed by functions $\mathrm{LayerNorm}$~\cite{ba_layer_2016} and $\mathrm{ReLU}$~\cite{nairRectifiedLinearUnits2010} (see Section~\ref{sec:appendix_arch}).

After the embedding, each node has a representation $v\in\mathbb{R}^{F}$. We assemble per-node representations into a single matrix $h^{(0)}\in\mathbb{R}^{N_\mathcal{G}\times F}$ that will be refined over $T$ iterations, with $N_\mathcal{G}=N_{\nD}+N_{\nEM}+N_{\nL}$ the total number of vertices in $\mathcal{G}$. 
Let $t$ be the $t$-th iteration, for $0\le t \le T$.
Each iteration applies a stack of $\nMP{=}2$ residual message-passing layers (Sect.~\ref{sec:residualmessagepassing}) followed by a GRU update
\begin{equation}
    h^{(t)} = \mathrm{GRU}\!\left(\mathrm{ResMP}^{\nMP}\!\left(h^{(t-1)},\,\mathcal{G}\right),\; h^{(t-1)}\right),
    \label{eq:gru_update}
\end{equation}
where $h^{(t)}$ is the graph representation at iteration $t$. Eq.~\eqref{eq:gru_update} constitutes a recursive definition for the $t$-th graph representation $h^{(t)}$.

The GRU output is passed through a learnable iteration embedding via Feature-wise Linear
Modulation (FiLM)~\cite{perezFiLMVisualReasoning2018}
\begin{equation}
    \tilde{h}^{(t)} = \sigma(\gamma_t)\odot h^{(t)} + \beta_t
\end{equation}
where $\odot$ denotes the element-wise (Hadamard) product, and $\gamma_t, \beta_t$ are produced from a learned
per-iteration embedding; $\beta_0$ is initialised to zero and $\sigma(\gamma_0)=1/2$.

A two-layer readout maps each node's final representation to a scalar logit. Over the error-mechanism and logical nodes the latter yield, at every iteration, patterns of fault mechanisms and logical value(s). The network is trained on the two supervised terms --- the true logical value(s) on the $\nL$-nodes and the true error pattern on the $\nEM$-nodes --- and they are combined across iterations with weights $w_t\propto t^2$ (see Eq.~\eqref{eq:loss}), so that the last iterations become more relevant.

The full architecture is illustrated in Fig.~\ref{fig:gnn_arch}; further details on
the message-passing layers are given in Appendix~\ref{sec:appendix_arch}. After the description of the model, we now outline its training.

\subsection{Training}\label{sec:training}

The decoder is trained on samples from \texttt{stim} that are always regenerated, so that there is no fixed data set: a fresh batch of shots is drawn at every optimisation step. No sample is ever reused, there are no epochs, and there is no notion of
overfitting a training set. Within each batch the physical error rate is drawn \emph{per shot}, log-uniformly in
an interval reaching from well below threshold up to $p \lesssim p_\mathrm{thr}$, so that a
single model covers the whole operating range at once. Shots whose syndrome is identically zero are
discarded, since at inference they would be handled exactly by predicting automatically no
logical flips.

We employ an exponential moving
average (EMA)~\cite{polyakAccelerationStochasticApproximation1992} to rebalance the weights of the models. Validation is run on the EMA against the classical baseline on the same shots --- correlated MWPM for the surface code. In this fashion, we are led to consider the mean of $\log(\mathrm{LER}_\mathrm{baseline}/\mathrm{LER}_\mathrm{GNN})$ as measure of the model quality. Training is distributed over $4$ GPUs with bfloat16 mixed precision~\cite{micikeviciusMixedPrecisionTraining2018,kalamkarStudyBFLOAT16Deep2019}. A
separate model is trained per code and distance, with the settings of
Table~\ref{tab:hyperparams}. The LaCross models $[\![34,4,3]\!]$--$[\![100,4,5]\!]$ and the first versions of the phenomenological surface-code models pre-date our current pipeline of the training and were obtained by training over a periodically regenerated data set --- a staged noise curriculum~\cite{bengioCurriculumLearning2009} was used for the surface code, while a fixed mixture of different noise rates was employed for LaCross. We note here this shift in the paradigm of the training for completeness' sake.

In the next section we treat the benchmark of our decoder against the state-of-the-art baseline for each code family. Furthermore, we discuss some aspects of post-selection and time of inference of the decoder.


\section{Results}\label{sec:results}

We devote this section to evaluating the decoder against the state-of-the-art baseline for each family --- correlated MWPM for the surface code and BP+OSD for the LaCross codes. The performance of the decoder is measured by the logical error rate (LER) as a function of the physical error rate $p$, with $r{=}d$ syndrome rounds, under \textit{phenomenological noise} for both families and additionally under \textit{circuit-level noise} for the surface code. We introduce the quantity
$\Lambda(p)~{=}~\mathrm{LER}_\mathrm{baseline}(p)/\mathrm{LER}_\mathrm{GNN}(p)$ as a measure of the advantage of the decoder against the baselines at a given physical error rate $p$. 
In the following it will be useful to consider also a quantity acting as weighted mean of $\Lambda(p)$ over different values of $p$. We introduce the symbol $\Lambda$ for a specific code as the geometric mean of $\Lambda(p)$
\begin{equation}
    \Lambda = \exp\!\Big(\tfrac{1}{|\mathcal{P}|}\textstyle\sum_{p\in\mathcal{P}}\log\Lambda(p)\Big),
    \label{eq:lambda_mean}
\end{equation}
over the set $\mathcal{P}$ of the noise points that lie below threshold. The use of this mean is justified since the logical error rate spans several orders of magnitude across the grid: the logarithmic average weighs every point equally and treats a gain and a loss of the same factor symmetrically. We stress that $\Lambda>1$ signals a LER lower than a baseline and therefore indicates a better performance. Table~\ref{tab:combined_results} lists $\Lambda$ for every code, and Fig.~\ref{fig:lambda_size} shows it as a function of the size of the decoding graph.

The logical error rates of all three settings are collected in Fig.~\ref{fig:hero_ler}. We first establish the LER on the surface code under phenomenological noise (Sec.~\ref{sec:sc_results}) and under circuit-level noise (Sec.~\ref{sec:circuit_results}), then on the LaCross codes (Sec.~\ref{sec:lacross_results}). 
In Sec.~\ref{sec:transfer} we introduce a ``generalist'' model, trained over a variety of different distances and families of codes. We then examine two properties beyond raw accuracy: decoding latency (Sec.~\ref{sec:time}) and confidence-based post-selection (Sec.~\ref{sec:postselection}).

Throughout, the architecture and the training procedure are the same; only the DEM changes.

\subsection{Surface code under phenomenological noise}\label{sec:sc_results}

The average decoder advantage $\Lambda$ over correlated MWPM ranges from $\Lambda\simeq1.19$ at $d=9$ to $\Lambda\simeq1.32$ at $d=7$, and is reported in Table~\ref{tab:combined_results} and Fig.~\ref{fig:lambda_size}. On the surface code the advantage has a peak at intermediate distances and decreases on the largest graphs (Fig.~\ref{fig:lambda_size}). We believe that monotonicity can be attained with further training.
The surface code provides a first benchmark for the GNN decoder. At any distance $d\in\{3,5,7,9\}$ and any phenomenological noise rate, our decoder attains a lower LER than MWPM and we plot its curve as a function of the noise in Fig.~\ref{fig:hero_sc}.

Our decoder satisfies a second metric that establishes its near-optimality as decoder: fitting the sub-threshold scaling of Eq.~\eqref{eq:ler_scaling} yields an exponent close to the ideal $(d+1)/2$ for every distance (cfr. Table~\ref{tab:combined_results}).

\subsection{Surface code under circuit-level noise}\label{sec:circuit_results}

In the circuit-level regime, to a more realistic and thus complex noise inside the physical circuit corresponds a more intricate graph-structure in the DEM. Namely, faults can propagate through the entangling gates and can flip three or four
detectors at once. This condition leads to a so-called non-graphlike structure --- as opposed to the phenomenological regime where the DEM has a graphlike structure. This entails that MWPM should operate a decomposition of the graph into graph-like parts, therefore partially losing its efficiency. As a consequence, to operate in this regime model, a correlated MWPM was devised and sets the classical baseline to be beaten by our GNN.

\begin{figure}[t]
\centering
\includegraphics[width=\linewidth]{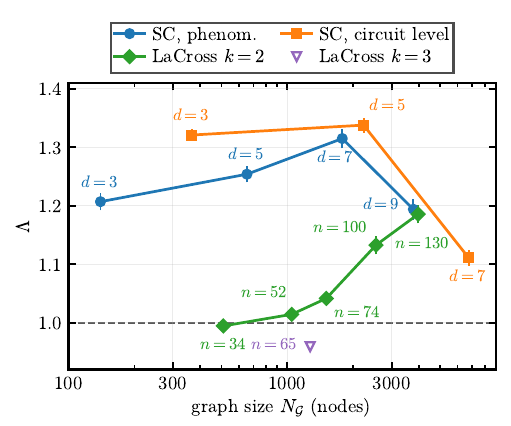}
\caption{Average advantage $\Lambda$ of the GNN over the reference baseline (Table~\ref{tab:combined_results}) as a function of the size of the graph (Table~\ref{tab:graph_size}). The reference baseline is correlated MWPM for the surface code (SC) and BP+OSD for LaCross; values above the show the advantage the GNN. Labels give the distance $d$ for the surface code and the number of physical qubits $n$ for LaCross.}
\label{fig:lambda_size}
\end{figure}

Against correlated MWPM the decoder makes fewer logical failures at every distance and every noise rate we probe (Fig.~\ref{fig:hero_circuit}): $\Lambda=1.32$ at $d=3$ and $1.34$ at $d=5$, i.e.\ about $25\%$ fewer failures, and $\Lambda=1.11$ at $d=7$ (Fig.~\ref{fig:lambda_size}).

We consider next high-rate LaCross codes, where MWPM is no longer available as benchmarking toolkit and we compare our decoder against BP+OSD.

\begin{table}[t]
\caption{Summary of GNN decoder performance.
    $\Lambda=\mathrm{LER}_\mathrm{baseline}/\mathrm{LER}_\mathrm{GNN}$, geometric mean over
    the sub-threshold points.
    Baseline: correlated MWPM for the surface code, BP+OSD for LaCross. Slope: fit of $\log\mathrm{LER}_\mathrm{GNN}$
    against $\log p$ over all points, to be compared with $(d+1)/2$ of
    Eq.~\eqref{eq:ler_scaling}.}
\label{tab:combined_results}
\begin{ruledtabular}
\begin{tabular}{llcc}
    Family & Code & $\Lambda$ & Slope \\
\hline
    \multirow{4}{*}{\shortstack[l]{Rot.\ SC\\(phenom.)}}
    & $d{=}3$ & $1.207(15)$ & 1.93 \\
    & $d{=}5$ & $1.254(14)$ & 3.05 \\
    & $d{=}7$ & $1.315(17)$ & 4.26 \\
    & $d{=}9$ & $1.194(18)$ & 5.51 \\[4pt]
    \multirow{3}{*}{\shortstack[l]{Rot.\ SC\\(circ.\ lev.)}}
    & $d{=}3$ & $1.321(10)$ & 1.97 \\
    & $d{=}5$ & $1.338(13)$ & 3.05 \\
    & $d{=}7$ & $1.111(14)$ & 4.18 \\[4pt]
    \multirow{5}{*}{\shortstack[l]{LaCross\\$k{=}2$}}
    & $[\![34,4,3]\!]$  & $0.994(6)$  & 2.15 \\
    & $[\![52,4,4]\!]$  & $1.014(6)$  & 2.41 \\
    & $[\![74,4,4]\!]$  & $1.042(6)$  & 2.61 \\
    & $[\![100,4,5]\!]$ & $1.133(16)$ & 3.65 \\
    & $[\![130,4,6]\!]$ & $1.186(16)$ & 4.69 \\[4pt]
    LaCross $k{=}3$ & $[\![65,9,4]\!]$ & $0.959(6)$ & 2.39 \\
    \end{tabular}
\end{ruledtabular}
\end{table}

\begin{figure}[t]
\centering
\includegraphics[width=\linewidth]{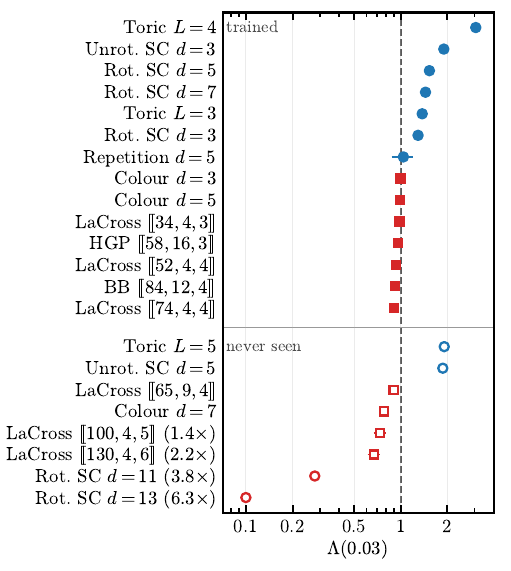}
\caption{A single set of weights across codes. Advantage
    $\Lambda(p=0.03)$ of the fifteen-code model at
    $p=0.03$; values above $1$ favour the GNN. Circles: against uncorrelated MWPM; squares: against BP+OSD. Blue: the GNN makes fewer logical errors than the
    baseline; red: more. Filled markers: codes in the training set; open
    markers: codes never presented to the model. The four largest unseen codes exceed the largest training graph by the factor indicated.}
\label{fig:universal}
\end{figure}
\subsection{LaCross code}\label{sec:lacross_results}

As for the surface code in the circuit level noise regime, also for LaCross codes the DEM contains hyperedges, making MWPM
inapplicable; we benchmark against BP+OSD~\cite{roffeDecodingAcrossQuantum2020,panteleevDegenerateQuantumLDPC2021}(Fig.~\ref{fig:hero_lacross}).
The primary results concern $k{=}2,\,3$ codes where, differently from the surface code, there
are several logical observables and a logical failure is an error over at least one of them.

On the $k{=}2$ family --- that has $k^2{=}4$ logical operators --- the advantage grows monotonically with the size of the code (Fig.~\ref{fig:lambda_size}). The
decoder is indistinguishable from BP+OSD on the smallest member
($\Lambda = 0.994 \pm 0.006$ on $[\![34,4,3]\!]$), slightly ahead on the next
($1.014 \pm 0.006$ on $[\![52,4,4]\!]$, $2.3\sigma$),
surpasses it on $[\![74,4,4]\!]$ ($\Lambda = 1.042 \pm 0.006$, $6.6\sigma$), and reaches
$\Lambda = 1.133 \pm 0.016$ on $[\![100,4,5]\!]$ and $\Lambda = 1.186 \pm 0.016$ on
$[\![130,4,6]\!]$ --- $16\%$ fewer logical failures than BP+OSD on the largest code of the
family, over $130$ physical qubits. Within each code the advantage likewise grows with the
noise, and the crossover falls around $p \approx 0.010$--$0.013$.

The trend does not, however, extend to a larger number of logical observables. On the
$k{=}3$ code $[\![65,9,4]\!]$, which encodes $k^2{=}9$ logical qubits,
the best model we have for that code remains below BP+OSD
($\Lambda = 0.959 \pm 0.006$). We interpret this negative result as the advantage in dealing with a bigger graph
having less impact than the difficulty of correcting nine logical observables at the same time.

Beyond accuracy, the GNN is also faster
per shot than BP+OSD on every code, most of all on the largest one --- precisely the regime
where BP+OSD becomes expensive --- as we detail in Sec.~\ref{sec:time}.

\subsection{One decoder for many codes}\label{sec:transfer}

While in the previous sections we considered models trained separately for each different code, distance and noise model, we discuss in this section a ``generalist'' model, naively trained over a range of different codes and distances, with a single set of trainable weights being thus shared across different DEMs. We trained one network --- $0.66$M parameters --- on the detection error models of fifteen codes, spanning
repetition, rotated and unrotated surface, toric, hypergraph-product, LaCross, bivariate-bicycle
and colour codes of up to $N_\mathcal{G}\sim1800$ nodes. Nothing in the network tells it which code it is decoding: the code is
known to it only through the DEM and its input features.

Fig.~\ref{fig:universal} illustrates a comparison between single GNN model and the classical baselines at $p=0.03$. On the graphlike codes, $1.3$--$3.1\times$ fewer logical failures were made than the classical baseline (uncorrelated MWPM) ($\Lambda(0.03)=1.04$--$3.05$) while, on the quantum codes carrying hyperedges, $\Lambda(0.03)=0.90$--$1.02$ when compared to BP+OSD. The same model also decodes quantum codes that were absent from its training: the unrotated surface code at $d=5$ and the toric code at $L=5$. Our decoder wins with about half the
logical failures, and then fails to generalize with $\Lambda(p)$ decreasing in the size of the decoded graph.

A few comments on the relevant figures of merit are in order.
\emph{(i) Size.} Accuracy degrades with the ratio between the size of the target graph and
that of the largest graph seen in training. \emph{(ii) Family.} Inside a homogeneous family, a model specifically trained on
its largest member dominates the generalist, so that the latter could be worth its cost only when the
target code is not known in advance. \emph{(iii) Noise model.} Transfer from phenomenological
to circuit-level noise fails completely, and fails already at $d=3$ for the rotated surface code, whose $366$-node graph lies
well inside the trained size range: this is due to the prior probabilities of the mechanisms falling below the whole range seen in
training and to a remarkably higher number of edges. This case was not reported in Fig.~\ref{fig:universal}.

Within a family, the practical route to reuse a model on a larger code is therefore not extrapolation
but initialisation. A surface-code decoder trained at $d=9$ and applied unchanged to $d=11$ is
$3.7\times$ worse than uncorrelated MWPM, and analogously for a LaCross model trained on $[\![100,4,5]\!]$ and tested on $[\![130,4,6]\!]$. Rather, retraining those models on the new codes gives a $\Lambda(p)=1.05$--$1.08$ against MWPM for $d=11$ rotated surface code above $p=0.03$, and $\Lambda(0.03)=1.186$ for $[\![130,4,6]\!]$ (cfr. Sec.~\ref{sec:lacross_results}).

In summary, this ``degenerate'' example of decoder highlights both the potential and the limitations of our approach to decoding in QEC. In a sense, a single decoder capable of performing the ``best possible'' correction for any code whose DEM is available would be the most desirable outcome of this research, since, for any stabilizer code, the DEM captures all the relevant information required by the decoder to perform error correction. In practice, however, a decoder trained specifically for a given code family and distance outperforms this general-purpose approach. We leave the investigation of possible improvements to the broader training strategy as an avenue for future work.

\subsection{Decoding time}\label{sec:time}

\begin{figure}[t]
    \centering
    \includegraphics[width=\linewidth]{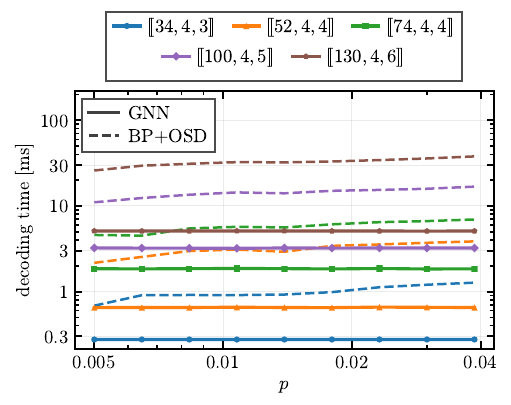}
    \caption{Effective per-shot decoding time $T_\mathrm{eff}$ versus the physical error rate $p$ for the $k{=}2$ LaCross codes, from
    $[\![34,4,3]\!]$ to $[\![130,4,6]\!]$, at $r{=}d$ rounds. Not only the GNN (solid) is faster: its latency is also
    flat in $p$, whereas for BP+OSD (dashed) it increases with it.}
    \label{fig:latency_E}
\end{figure}

Accuracy is not the only requirement for a decoder to be useful: it should also be fast enough to keep up with the syndrome stream. As PyMatching, the reference implementation of MWPM, for the surface code is already greatly optimized both in accuracy and in efficiency, we compare instead the decoding time of our GNN against BP+OSD. Two aspects of the decoding time deserve to be distinguished: a per-shot global cost that is amortized over an entire batch, and the true single-shot latency, which is the relevant quantity for real-time decoding. For reference, we report that BP+OSD runs with the following standard parameters: min-sum with scaling factor $0.4$, at most $100$ iterations, and combination-sweep OSD of order $10$. Both decoders are timed on the same circuits with
$r{=}d$ syndrome rounds. We record the time of each decoder on its natural platform: the GNN on a single NVIDIA H200 GPU, BP+OSD on a single core of an Intel Xeon Gold 6226R CPU with the \texttt{ldpc} library~\cite{roffeLDPCPythonTools2022,roffeDecodingAcrossQuantum2020}. The comparison is therefore carried on different hardware; what does not depend on the platform is that the cost of the GNN is independent of the noise, while that of BP+OSD grows with it.

In order to characterize the cost in time, we decompose the decoding time of a batch of $B$ syndromes as $T_\mathrm{lat}(B)=T_\mathrm{oh}+T_\mathrm{eff}\cdot B$, so that the per-shot latency be $T_\mathrm{oh}/B+T_\mathrm{eff}$. Crucially, $T_\mathrm{oh}$ is a fixed per-invocation overhead --- therefore amortized over bigger batches --- and $T_\mathrm{eff}$ is the effective per-shot decoding cost. 

Plotting $T_\mathrm{eff}$ reveals the difference in nature between the two decoders
(Fig.~\ref{fig:latency_E}). For the GNN, $T_\mathrm{eff}$ is flat in the physical error rate. The network performs the same fixed amount of arithmetic whatever the syndrome, so its cost is set by the size of graph and of the neural architecture, not by the noise. Conversely, $T_\mathrm{eff}$ for BP+OSD grows with $p$ across the family over the range shown,
since denser syndromes trigger more belief-propagation and ordered-statistics
work~\cite{roffeDecodingAcrossQuantum2020}.

As a result the GNN is cheaper per shot on every code of the family, by a factor between $2.5$ and
$5.1$ at the lowest noise and between $3.8$ and $7.4$ at the highest, with the margin that widens with $p$. It is widest on the largest code, $[\![130,4,6]\!]$ ($5.1$--$7.4\times$), where a
single BP+OSD decode already takes $26$--$38$~ms against $5.1$~ms for the network.

It is important to point out that $T_\mathrm{oh}$ determines which regime is relevant. For
BP+OSD it is negligible ($8$--$56\,\mu$s), whereas for the GNN it is not ($0.3$--$1.2$~ms): it is
the cost of launching the computation. However, on larger codes and bigger batches $T_\mathrm{oh}$ is readily
amortised.

We have therefore also measured the strict single-shot regime $B=1$, where $T_\mathrm{oh}$ is not
amortised at all and the comparison is least favourable to the GNN. At $p\simeq0.018$ the decoder
takes $1.9$~ms on $[\![52,4,4]\!]$, $3.3$~ms on $[\![74,4,4]\!]$, $4.7$~ms on $[\![100,4,5]\!]$ and
$2.0$~ms on $[\![65,9,4]\!]$ (the last one is not shown in the picture), that is $1.7$, $1.9$, $3.2$ and $2.0$ times less than single-threaded
BP+OSD, whereas on the smallest code $[\![34,4,3]\!]$ BP+OSD remains $1.3\times$ faster. 

As a consequence we conclude that the real-time advantage of the GNN exists but it is not ubiquitous, and rather it is established above a code size and then grows from there. Remarkably, this is the same direction in which the advantage in accuracy grows (Fig.~\ref{fig:lambda_size}).

\textit{Towards real-time decoding.} Real-time decoding has two main prerequisites. 
Firstly, the decoder should keep pace with the stream of syndromes output through the computation --- each round processed faster than it is produced, lest the decoder fall into the so-called \textit{backlog problem}~\cite{terhal2015}.
Secondly, its response should be returned within the reaction time of the hardware. Per syndrome round, the single-shot latencies above correspond to $0.5$--$0.9$~ms, already comparable to the time-scales of neutral-atom processors~\cite{bluvstein2024}, the natural platform for LaCross codes. 
Our architecture makes this regime approachable. Indeed, the computation is a fixed sequence of operations on a never-changing graph, without the data-dependent pipeline of ordered-statistics decoding in BP+OSD: its latency is deterministic and well suited to dedicated hardware. 
Furthermore, already existing streaming decoders rely on overlapping windows of rounds~\cite{skoric2023,tan2023}, and the detection error model of a window is itself a DEM: the same network applies unchanged, trained on window DEMs. We leave both directions to future work.

\subsection{Post-selection}\label{sec:postselection}

Finally, in this paragraph we investigate how the probabilistic output of the GNN can be leveraged for soft decisions~\cite{bauschLearningHighaccuracyError2024}. To each shot we assign a \textit{confidence} $c$ defined as
\begin{equation}\label{eq:confidence}c = \min_{k=1,\dots,N_{\nL}}\max( \hat{p}_k,\,1-\hat{p}_k),\end{equation} 
where $\hat{p}_k$ is the marginal of the $k-$th logical observable computed by the GNN. Intuitively, it quantifies how far the GNN is from being certain that a logical observable $\ell$ has happened ($p_\ell=1$) or not ($p_\ell=0$), so that a value of $p_\ell=1/2$ would mean that the GNN is maximally undecided with respect to such observable. By definition, the confidence is the marginal of the least certain logical
observable. It is possible to devise a strategy to keep or discard single shots in the following fashion. 
Fix a threshold $\tau$ for the confidence; a shot is to be accepted if $c \ge \tau$ and else rejected. 
Raising
$\tau$ discards the least-confident shots (Fig.~\ref{fig:postselection}, top row
of each panel): on the surface code the accepted LER drops by by a factor $14$--$40$ at $\tau=0.99$ --- while keeping between $67\%$ ($d=3$) and $99.6\%$ ($d=9$) of the shots. It is remarkable but not surprising that the discarded shots are not random: rejection concentrates on the high-weight error events (Fig.~\ref{fig:postselection}, bottom row). A distance-$d$ code corrects all errors of weight up to $\lfloor(d-1)/2\rfloor$, while a logical failure requires an error of weight at least $d$, the minimum weight of a logical operator (dotted line); post-selection preferentially rejects precisely the high-weight events, which lie in the regime where the decoder can fail. On platforms where the logical state can be re-prepared, such as neutral-atom processors~\cite{pecorariAddressableGatebasedLogical2025}, the low-confidence shots can be flagged for re-measurement, lowering the effective logical error rate at a small cost.

\section{Discussion and Outlook}\label{sec:outlook}

We have introduced a graph neural network decoder that works directly with the detection error model, which is all that an optimal decoder should see in order to operate its correction. What makes special the GNN is a code-agnostic structure, that makes it applicable to any stabilizer code that can be compiled into a DEM. The GNN decoder
outperforms correlated MWPM on the rotated surface code at every distance up to $d=9$ under phenomenological
noise and up to $d=7$ under circuit-level noise, and matches or
surpasses BP+OSD on the LaCross $k{=}2$ family, with an advantage that grows with the size of
the code (Fig.~\ref{fig:lambda_size}). Its cost per shot does not depend
on the physical error rate, which makes it cheaper than BP+OSD, and its probabilistic output enables the possibility to operate a confidence-based post-selection. Moving from one
code, family or noise model to another required no change to the network, only a new DEM and a
new training run --- which, in the streaming regime of Sec.~\ref{sec:training}, takes from hours to a few days on four GPUs.

\textit{Outlook.} Three avenues follow in a natural way. First, the generalist model deserves a
more systematic study: its accuracy is limited by the ratio between the target graph and the
largest training graph rather than by the number of codes or of parameters
(Sec.~\ref{sec:transfer}), so that it would be interesting to train on larger graphs, mix noise models in the
training set, test on an entire family held out from training, and use the generalist as
an initialisation for dedicated models. Secondly, since the decoder only requires a DEM, other
error models can be addressed without modification: atom loss, where the DEM changes from shot to
shot~\cite{perrinQuantumErrorCorrection2025, perrinCorrelatedAtomLoss2026}, erasure-biased noise~\cite{wuErasureConversionFaulttolerant2022},
leakage and biased noise, as well as logical operations such as lattice
surgery~\cite{horsmanSurfaceCodeQuantum2012}, towards which neural decoders have recently been
extended from our regime of memory experiments~\cite{bonillaataidesNeuralDecodersUniversal2025}; training or
fine-tuning on experimental syndromes~\cite{bauschLearningHighaccuracyError2024} is the
corresponding step towards hardware.
Finally, the fixed, syndrome-independent cost of the network
makes it a great candidate for dedicated hardware, as FPGA implementations of GNN decoders already show~\cite{ciceroLowLatencyGNN2026} so to turn the measured advantage into a real-time one.


\section*{Code and data availability}
The source code of POLYMECHANON --- construction of the tripartite graph and of its features from
the \texttt{stim} detection error model, the network, and the streaming training pipeline ---
together with the benchmarking scripts, the trained models and the data underlying all figures
and tables is available from the authors upon reasonable request.

\section*{Acknowledgements}

We are grateful to Hugo Perrin for his thoughtful feedback and for his ideas that prompted us in first place to explore this line of research.
This research has received funding from the European Union's Horizon Europe Research
and Innovation Programme under the Marie Sk\l{}odowska-Curie GA number 10120240 (MLQ),
 and
from the French National Research Agency under the Investments of the Future Program
projects ANR-21-ESRE-0032 (aQCess), ANR-22-CE47-0013-02 (CLIMAQS),
ANR-17-EURE-0024 (QMat), and ANR-22-CMAS-0001 France 2030 (QuanTEdu-France).
We thank QPerfect for providing part of the computing resources used in this work.
The authors would like to acknowledge the High Performance Computing Center of the University of Strasbourg for supporting this work by providing scientific support and access to computing resources. Part of the computing resources were funded by the Equipex Equip@Meso project (Programme Investissements d'Avenir) and the CPER Alsacalcul/Big Data.

\appendix

\section{Circuit construction and DEM extraction}\label{sec:appendix_stim}

Circuits are generated using \texttt{stim}~\cite{gidneyStimFastStabilizer2021}. A memory-$Z$ experiment
consists of three stages: (\textit{i})~initialisation of all data qubits in $|0\rangle$;
(\textit{ii})~$r$ rounds of stabilizer measurement; (\textit{iii})~final data-qubit
readout. How measurements are performed depends on the code itself --- via ancilla qubits and CZ gates for the surface code and via joint Pauli-product measurement for LaCross code. Anyhow, the decoder only operates on the DEM and is insensitive to different measurement procedures.
Detectors are defined as XOR of consecutive measurements
of the same stabilizer: 
\begin{equation}
    d_{i,r} = s_{i,r} \oplus s_{i,r-1},
\end{equation}
where $s_{i,r}$ is the outcome of the $i-$th stabilizer at round $r$. Since at the beginning (respectively end) there is no previous (subsequent) round, the first-round  (last-round) detectors are defined by confronting the first (last) syndrome with the known initial (final)  values coming from the initialization (\textit{i}) (respectively readout (\textit{iii})).
The logical observables are the code's logical $Z$-operators evaluated on the final data measurements: a single observable for the surface code and $k^2\,{=}\,4,\,9$ for the LaCross code.

In the phenomenological noise regime, during each round a single-qubit depolarizing noise is applied to the data qubits together with an independent bit-flip error to every stabilizer measurement, both with probability $p$ (Table~\ref{tab:noise_channels}). In the circuit-level regime, used for the surface code in Sec.~\ref{sec:circuit_results}, noise is additionally applied to the ancillas, after every reset and after every entangling gate, all at the same rate $p$; it is this last channel that produces correlated faults that may flip more than two detectors. The code-capacity regime is listed for reference but is not used here.

\begin{table}[ht]
\caption{Noise channels per model. $p$ = physical error rate.}
\label{tab:noise_channels}
\begin{ruledtabular}
\begin{tabular}{p{3.9cm}ccc}
\rl Channel & \shortstack{Code\\cap.} & \shortstack{Phenom.\\\mbox{}} & \shortstack{Circuit\\lev.} \\
\hline
\rl Data qubit depolarization      & $p$ & $p$ & $p$ \\
\rl Ancilla qubit depolarization    & --- & --- & $p$ \\
\rl Measurement flip                & --- & $p$ & $p$ \\
\rl Reset flip                      & --- & --- & $p$ \\
\rl After-gate $2$-qubit depolarization & --- & --- & $p$ \\
\end{tabular}
\end{ruledtabular}
\end{table}

Within the DEM, error mechanisms
with identical detector and logical support might be present. When this is the case, they are merged with probabilities combined as
$p_\mathrm{merged} = p_1(1{-}p_2) + p_2(1{-}p_1)$.

\section{GNN architecture}\label{sec:appendix_arch}

We recall notations: $F$ is the number of per-node hidden features and $T$ the
number of iterations. All models use $\nMP=2$ residual message-passing layers per iteration. $\mathcal{N}(i)$ denotes the graph neighbours of a node $i=1,\dots,N_\mathcal{G}$ in the graph representation $\mathcal{G}$ of the DEM, while $\mathrm{LN}$ is LayerNorm \cite{ba_layer_2016}, and $[\,\cdot\,\|\,\cdot\,]$ is the concatenation. In the following, we put the accent on architectural parts of the GNN that we have only briefly introduced in the previous sections.

\subsection{Subtype embedders}\label{sec:subtype_embedders}

Each of the six subtypes $s\in\{0,\ldots,5\}$ has an independent linear embedder $\phi_s: \mathbb{R}^9 \to \mathbb{R}^F$ defined by $x_i\mapsto W_s\, x_i + b_s$ --- for $W_s \in \mathbb{R}^{F\times 9}$, $b_s\in\mathbb{R}^F$ --- and followed by the non-linear functions $\mathrm{LN}$ and $\mathrm{ReLU}$; together they is used to define a per-node representation
\begin{equation}
    h_i^{(0)} = \mathrm{ReLU}\!\left(\mathrm{LN}\!\left(\phi_s(x_i)\right)\right)\in\mathbb{R}^F,
    \quad s = \mathrm{subtype}(i).
\end{equation}
The nine-dimensional inputs of a node $x_i$ are listed in Table~\ref{tab:features}, where the caption describes each entry. 
Concatenation $h^{(0)}=[h^{(0)}_1||\dots||h^{(0)}_{N_\mathcal{G}}]$ gives the initial graph representation that is to be updated as prescribed in Eq.~\eqref{eq:gru_update}. We detail its specifics in the next paragraph.

\subsection{Residual message passing}\label{sec:residualmessagepassing}

Recall that in Eq.~\eqref{eq:gru_update}, $\mathrm{ResMP}^{\nMP}$ is the composition of
$\nMP$ identical residual layers. In the following we describe how; the resulting output is the candidate state that the GRU combines with $h^{(t-1)}$ --- i.e. the graph representation at iteration $t-1$.

During each of the $T$ iterations, each of the $\nMP$ residual layers computes edge messages and updates node states:
\begin{align}
    \bar h_i^{(t)} &= \mathrm{LN}(h_i^{(t)}), \\
    m_{ij}^{(t)} &= \alpha_{ij}\,\mathrm{MLP}_e\!\left([\bar h_i^{(t)} \,\|\, \bar h_j^{(t)}]\right) \\
    h_i^{(t)} &\mathrel{+}= \sum_{j\in\mathcal{N}(i)} m_{ij}^{(t)}
\end{align}
where $h_i^{(t)}$ is the per-node representation at iteration $t$, the subscript $e$ stands for \emph{edge} and $\mathrm{MLP}_e$ is a two-layer perceptron~\cite{goodfellowDeepLearning2016} defined by $\mathrm{MLP}(x)=W_2\,\mathrm{ReLU}(W_1x+b_1)+b_2$. The perceptron, after concatenation, turns the pair of ``neighboring'' per-node representations into a message
$m_{ij}\in\mathbb{R}^F$ to be sent along the edge $(i,j)$; the message is further scaled by
$\alpha_{ij}=\mathrm{sigmoid}(a^\top[\bar h_i^{(t)}\|\bar h_j^{(t)}])$ --- with $a\in\mathbb{R}^{2F}$ a learnable vector --- and finally aggregated and summed to $h_i^{(t)}$.
This is followed by a two-layer feed-forward network (FFN)~\cite{vaswaniAttentionAllYou2017}: $h_i \mathrel{+}= \mathrm{FFN}(\mathrm{LN}(h_i^{(t)}))$. 

\subsection{Output head}\label{sec:output_head}

At each iteration $t=1,\dots,T$, the logical prediction is read from the $\nL$-nodes:
\begin{equation}
    \hat{\ell}^{(t)} = \mathrm{MLP}_\mathrm{out}\!\left(\tilde{h}_{\nL}^{(t)}\right)
    \in \mathbb{R}
\end{equation}
where $\mathrm{MLP}_\mathrm{out}$ is a two-layer network with ReLU activation that produces one logit per logical observable; the subscript \emph{out} marks it as the output (readout) network, as opposed to the message network $\mathrm{MLP}_e$.

\subsection{Training loss}\label{sec:training_loss}
At iteration $t$ the shared readout produces a predicted error-pattern $\hat{e}^{(t)}$ and
logical error $\hat{\ell}^{(t)}$ by outputting their probabilities on the $\nEM$ and $\nL$ nodes respectively. The
per-iteration loss is the sum of the two supervised --- i.e. computed against the true error pattern and logical flip --- binary cross-entropy~\cite{goodfellowDeepLearning2016},
\begin{equation}
    \mathcal{L}^{(t)} = \delta\,\mathrm{BCE}(\hat{\ell}^{(t)}\!,\ell)
      + \varepsilon\,\mathrm{BCE}(\hat{e}^{(t)}\!,e),
    \label{eq:loss_iter}
\end{equation}
where $(e,\ell)$ are the ground-truth error mechanisms and logical observables and
$(\delta,\varepsilon)$ are fixed weights. 
This is the function minimised during training: its gradient computed with respect to the network parameters is derived by backpropagation~\cite{rumelhartLearningRepresentationsBackpropagating1986,goodfellowDeepLearning2016} and employed to update them with the AdamW optimiser~\cite{kingmaAdamMethodStochastic2015,loshchilovDecoupledWeightDecay2019}.

For codes with more than one logical observable the
first term is the mean over the number of observables. The per-iteration losses are aggregated with quadratic weights
that emphasise the later iterations:
\begin{equation}
    \mathcal{L} = \sum_{t=1}^{T} w_t\,\mathcal{L}^{(t)},
    \qquad w_t = \frac{t^2}{\sum_{s} s^2}.
    \label{eq:loss}
\end{equation}

An earlier form of the objective carried three additional terms: a supervised cross-entropy
on the syndrome, $\beta\,\mathrm{BCE}(\hat{\sigma}^{(t)},\sigma)$, and two consistency terms
$\alpha\,\rho(H\hat{e}^{(t)},\hat{\sigma}^{(t)})$ and
$\gamma\,\rho(L\hat{e}^{(t)},\hat{\ell}^{(t)})$, with $\rho$ the smooth-$\ell_1$ loss~\cite{girshickFastRCNN2015} and
$H,L$ the matrices of Sec.~\ref{sec:dem}, with the reduction modulo $2$ replaced by the function $x\mapsto|\sin(\pi x/2)|$ --- which is a differentiable substitute of the function $x \bmod 2$ defined on the integers. Requiring the predicted error pattern to reproduce the syndrome and the
predicted logical is an appealing way of using the code structure inside the training loss,
but we found no statistical evidence of a better convergence and therefore we limited ourselves to considering Eq.~\eqref{eq:loss} as training loss.

\bibliography{QML_updated}

\begin{figure*}[tp]
    \centering
    \begin{subfigure}[b]{\textwidth}
        \centering
        \includegraphics[width=\linewidth]{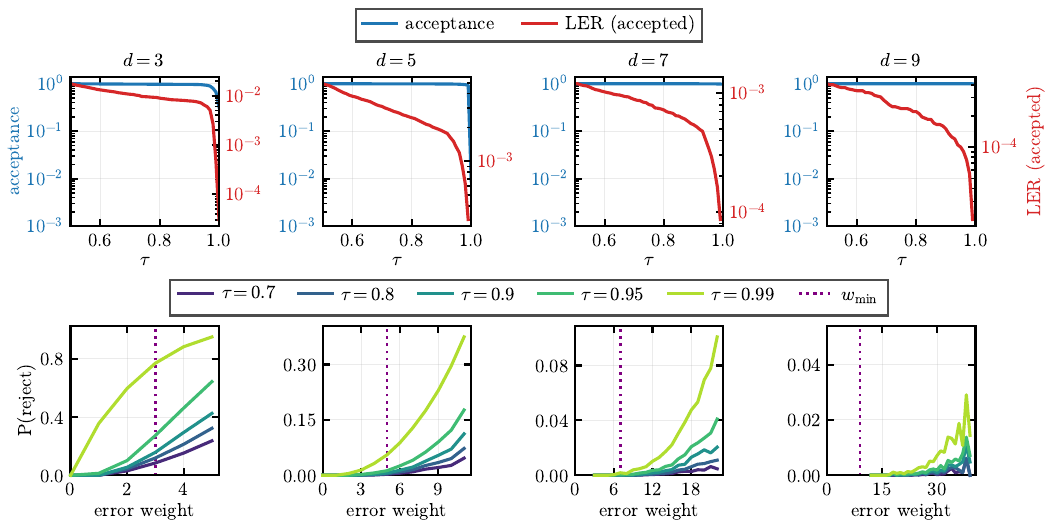}
        \caption{Rotated surface code, $d\in\{3,5,7,9\}$.}
        \label{fig:postselect_sc}
    \end{subfigure}

    \vspace{0.6em}

    \begin{subfigure}[b]{\textwidth}
        \centering
        \includegraphics[width=\linewidth]{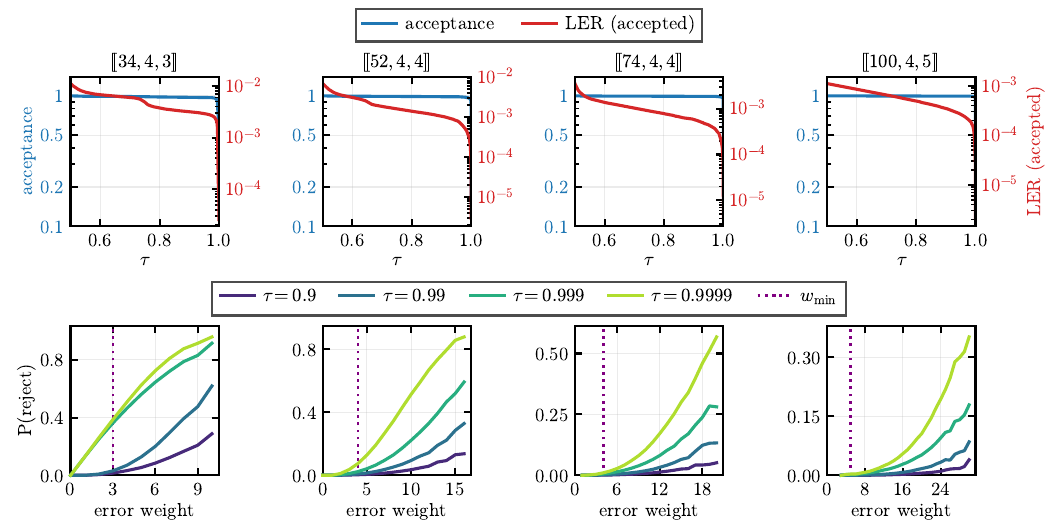}
        \caption{LaCross $k{=}2$ codes: $[\![34,4,3]\!]$, $[\![52,4,4]\!]$, $[\![74,4,4]\!]$ and $[\![100,4,5]\!]$.}
        \label{fig:postselect_lacross}
    \end{subfigure}

    \caption{Confidence-based post-selection.
    \subref{fig:postselect_sc}~surface code, \subref{fig:postselect_lacross}~LaCross.
    \emph{Top of each panel}: fraction of accepted shots (blue, log) and logical error rate --- computed at fixed error probability $p=0.015$ --- on the accepted shots (red, log) versus the confidence threshold $\tau$.
    \emph{Bottom}: rejection probability versus error weight for several $\tau$; rejection concentrates on errors with high weight. The dotted line marks the code distance $d$, i.e. the minimum weight of a logical operator (a distance-$d$ code corrects all errors of weight up to $\lfloor(d-1)/2\rfloor$). For the surface code at $\tau=0.99$ the accepted fraction is $67\%$ at $d=3$ and above $93\%$ for $d\ge5$, while the LER on the accepted shots is $14$--$40$ times lower than without post-selection; error weights with fewer than $200$ shots are not shown.
    \label{fig:postselection}
    }
\end{figure*}

\end{document}